\documentclass[epj]{svjour}
\usepackage{graphics}
\newcommand{\be}{\begin{equation}}
\newcommand{\ee}{\end{equation}}
\newcommand{\ba}{\begin{eqnarray}}
\newcommand{\ea}{\end{eqnarray}}

\newcommand{\im}{\mbox{Im}\,}

\newcommand{\Od}{{\cal O}}
\newcommand{\lsim}{\raise.3ex\hbox{$<$\kern-.75em\lower1ex\hbox{$\sim$}}}

\begin{document}
\title{Chiral Symmetry and light resonances in hot and dense matter}
\author{D.~Cabrera \and D.~Fern\'andez-Fraile
\and A.~G\'omez Nicola\thanks{Electronic address: gomez@fis.ucm.es}% etc
}                     % Do not remove
\institute{Departamento de F\'{\i}sica Te\'orica II. Univ.
Complutense. 28040 Madrid. Spain.}
\date{Received: date / Revised version: date}
\abstract{
We present a study of the $\pi\pi$ scattering amplitude in the $\sigma$ and
$\rho$ channels at finite temperature and nuclear density within a chiral
unitary framework. Meson resonances are dynamically generated in our approach,
which allows us to analyze the behavior of their associated scattering poles
when the system is driven towards chiral symmetry restoration. Medium effects
are incorporated in three ways: (a) by thermal corrections of the unitarized
scattering amplitudes, (b) by finite nuclear density effects associated to a
renormalization of the pion decay constant, and complementarily (c) by
extending our calculation of the scalar-isoscalar channel to account for finite
nuclear density and temperature effects in a microscopic many-body
implementation of pion dynamics. Our results are discussed in connection with
several phenomenological aspects relevant for nuclear matter and Heavy-Ion
Collision experiments, such as $\rho$ mass scaling vs broadening from dilepton
spectra and chiral restoration signals in the $\sigma$ channel. We also
elaborate on the molecular nature of $\pi\pi$ resonances.
 % end of PACS codes
} %end of abstract
\maketitle
\section{Introduction.}
\label{sec:intro}

The lightest meson resonances, the $\rho(770)$ and the $f_0(600)$ or $\sigma$,
play a crucial role in different phenomena pertaining to the hot and/or dense
medium created both in Relativistic Heavy Ion Collisions and in Nuclear Matter
experiments. The modifications of the spectral function of the $\rho$ resonance
in medium are crucial to understand correctly the dilepton yield emerged from
Heavy Ion Collisions \cite{CERES,NA60}. The two main theoretical scenarios
currently proposed in the literature can be classified into resonance mass
shifting and broadening, according to the resulting dominant effect. Mass
shifting models are inspired by the Brown-Rho (BR) scaling hypothesis
\cite{brownrho91}, which predicted that vector meson masses should scale with
the quark condensate and therefore the main spectral modification of the
resonance would be dictated by chiral symmetry restoration. This scenario is
supported also by the so called hidden local symmetry approach \cite{harada}.
The broadening-dominated scenario is supported by different theoretical analyses
\cite{Herrmann:1992kn,Pisarski:1995xu,Peters:1997va,Rapp:1997fs,Rapp:1999ej}
including recent Unitarized Chiral approaches
\cite{Cabrera:2000dx,Dobado:2002xf,FernandezFraile:2007fv}. The most recent
experimental dimuon data from the NA60 Collaboration \cite{NA60} clearly favor a
broadening situation with negligible mass shift, whereas the earlier CERES
results \cite{CERES} were reasonably explained by both descriptions. It is worth
mentioning also the results of the STAR Collaboration at RHIC \cite{STAR}, which
has reported a sizable mass reduction by medium effects measured in
$\rho^0\rightarrow\pi^+\pi^-$ instead of dileptons. The modifications of the
$\rho$ properties have been also measured in cold nuclear matter experiments.
The E325-KEK collaboration \cite{Naruki:2005kd} has reported a measurable shift
in the masses of vector mesons compatible with theoretical predictions based on
Brown-Rho scaling \cite{br02} and QCD sum rules \cite{Hatsuda:1995dy}. On the
other hand,  the JLab-CLAS experiment \cite{:2007mga}  has obtained results
compatible with vanishing mass shift, as predicted by most of in-medium hadronic
many-body analyses where broadening is the dominant effect
\cite{Herrmann:1992kn,Peters:1997va,Cabrera:2000dx,Urban:1998eg}.

The possible modification of the $f_0(600)/\sigma$ in hot and dense matter is
interesting because this is a state with the same quantum numbers as the vacuum
and therefore it might be sensitive to chiral symmetry restoration. In this
sense, an early proposal \cite{hatku85} suggested that the $\sigma$ could induce
a measurable threshold enhancement of the $\pi\pi$ cross section, which would be
interpreted as a precursor of chiral symmetry restoration. The argument was that
the mass of the $\sigma$ state should decrease by medium effects, since it is
proportional to its vacuum expectation value in the chiral limit. Such a
decrease would eventually shrink the available two-pion phase space when the
$\sigma$ mass reaches the two-pion threshold, producing a bump in the imaginary
part of the scattering amplitude, due to the proximity of the pole to the real
axis. It is important to remark that in this original argument, it is implicitly
assumed that i) the $\sigma$ is dominated by its $\bar q q$ component so that
its expectation value behaves like the quark condensate and ii) that the
$\sigma$ is narrow enough so that its width (imaginary part of the pole)
vanishes when its mass (real part) approaches the threshold. None of these
assumptions seem to be supported by the physical (vacuum) $f_0(600)$ state
quoted by the Particle Data Group \cite{Yao:2006px}, which is a very broad state
measured in $\pi\pi$ scattering. Although it is commonly accepted that this
state is a member of the scalar nonet, its $\bar q q$ nature has been criticized
on the basis of lattice \cite{Alford:2000mm} and large-$N_c$
\cite{Pelaez:2003dy} analyses. The physical state is likely to have important
non-$\bar q q $ compound  such as tetraquark, glueball or meson-meson,  commonly
referred to as a ``molecular'' state. Nevertheless, threshold enhancement is
indeed observed in nuclear matter experiments, both in $\pi A\rightarrow \pi\pi
A'$ \cite{Bonutti:2000bv,cb} and in $\gamma A\rightarrow \pi\pi A'$
\cite{messetal} reactions. Although the size of the effect is still under
debate, a clear signal is seen in the scalar channel for increasing nuclear
density as compared with the vector channel and this is in fair agreement with
most theoretical analyses at finite nuclear density
\cite{Chiang:1997di,davesne00,jihatku01,roca02,patkos03,cab05}. In contrast,
finite temperature analyses show that this state remains broad even near the
chiral phase transition despite the proximity of the pole to the
 two-pion threshold
\cite{FernandezFraile:2007fv,patkos03,hidaka04}, which in practice does not
produce any sizable enhancement in the scattering amplitude or cross section.
A striking possibility to be explored
is that  by increasing further the medium strength, the $\sigma$ could become a
$\pi\pi$ bound state, as suggested earlier in \cite{Schuck:1988jn} and confirmed
recently in \cite{FernandezFraile:2007fv,patkos03,hidaka04}.

In the present work, we will investigate further about these issues,
within the context of unitarized chiral approaches. The main goal is
to establish to what extent  chiral symmetry dictates  the in-medium
properties of the light meson resonances. The most general framework
to account for all the interactions compatible with chiral symmetry
is the effective chiral lagrangian approach. The most prominent
example is Chiral Perturbation Theory (ChPT) for the meson sector
\cite{we79,Gasser:1983yg}, but it can be equally applied to the
meson-baryon one \cite{Bernard:1995dp}. Since these effective
theories are built basically as expansions in derivatives or
energies, they cannot account for resonances, since the chiral expansion
violates the unitarity bounds. This has
been traditionally solved by introducing unitarization methods,
giving rise to the so called chiral unitary approaches, which have
proved to be very successful in vacuum to describe meson-meson and
meson-baryon interactions and generate dynamically low lying
resonances \cite{unitvaciam,unitvac}. Furthermore, as commented
above, the unitarization program has been extended to account for
finite temperature and density effects
\cite{Cabrera:2000dx,Dobado:2002xf,FernandezFraile:2007fv,roca02,cab05,Ramos:1999ku,Tolos:2008di}.

Here we will analyze some of our recent results for the $\rho$ and
$\sigma$ mesons obtained within the unitarized chiral framework,
paying special attention to their nature and their role in chiral
symmetry restoration and studying some aspects not considered before
like a new analysis of the combined effects of temperature and nuclear
density for the $\sigma$ meson based on a Lippmann-Schwinger (or Bethe-Salpeter)
equation
approach accounting for many-body pion dynamics
versus a simplified $f_\pi$-scaling scenario, a
direct comparison with nuclear matter experiments of our results for
the $\rho$ mass linear density dependence and the interpretation of
the results obtained for the behavior of the resonances near
threshold in terms of a ``molecular'' classification of those states.

The paper is organized as follows: in Section \ref{sec:finitetemp}
we will present the formalism and results within the framework of
the Inverse Amplitude Method (IAM) at finite temperature. In Section
\ref{sec:finitedenTzero} we will introduce nuclear density effects
only by rescaling properly the pion decay constant at $T=0$ within
the IAM. In that section, we will provide in particular an
interpretation of our results in terms of ``molecular''
classification and a numerical comparison with experimental results from
dilepton decays in resonance production in finite nuclei.
Finally, in Section \ref{sec:finitedentemp} we
present a new calculation for the $\sigma$ channel which includes
temperature and nuclear density many-body effects in order to
compare our different approaches.

\section{Finite temperature resonances with the Inverse Amplitude Method.}
\label{sec:finitetemp}

One of the simplest and more powerful unitarization methods for
chiral theories is the so called Inverse Amplitude Method (IAM)
\cite{unitvaciam}. Its name comes from the simple observation that
unitarity implies that the inverse of a given partial wave amplitude $t^{IJ}$
in $\pi\pi\rightarrow\pi\pi$ scattering should satisfy:

\ba S^\dagger S=1&\Rightarrow& \im t^{IJ}(s)=\sigma_0(s)\vert
t^{IJ}(s) \vert^2 \nonumber \\ &\Rightarrow&
\im\frac{1}{t^{IJ}(s)}=-\sigma_0(s) \label{unit}\ea for
$s>4m_\pi^2$,where $s$ is the center of mass energy squared and
$\sigma_0(s)=\sqrt{1-4m_\pi^2/s}$ is the two-pion phase space.

Consider now the ChPT expansion of partial wave amplitudes:

\be t^{IJ}(s)=t^{IJ}_2(s) + t^{IJ}_4(s) +
\Od(p^6)\label{pertunit}\ee

Here,  $p$ denotes generically a meson momentum, mass or temperature
($p$ is to be compared with the characteristic chiral scale
$\Lambda_\chi\sim$ 1 GeV, whereas $T$ is meant to be below
$T_c\simeq$ 200 MeV) and $t_k$ is the $\Od(p^k)$ contribution.
Recall that, according to the standard ChPT power counting
\cite{we79,Gasser:1983yg}, $t_2$ accounts for tree level diagrams
from the lowest order lagrangian ${\cal L}_2$. Up to that order,
only the pion decay constant $f_\pi$ and the pion mass $m_\pi$ enter
the result. The order $t_4$ includes the one-loop diagrams from
${\cal L}_2$ plus the tree level ${\cal L}_4$ terms needed for
renormalization. The ${\cal L}_4$  low-energy constants entering the
pion scattering amplitude, when it is expressed in terms of the
physical $m_\pi$, $f_\pi$ are denoted $\bar l_1-\bar l_4$ in the
convention of \cite{Gasser:1983yg}.

The chiral expansion (\ref{pertunit}) satisfies only a perturbative
version of the unitarity relation (\ref{unit}), namely:

\be \im t^{IJ}_4(s)=\sigma_0(s)\vert t_2^{IJ}(s) \vert^2 \ee and so
on for higher orders, which eventually means that chiral expansions
are not compatible with the bounds on partial waves implied by
unitarity. In other words, they grow arbitrarily with energy.
Unitarization methods allow to construct chiral amplitudes that are
exactly unitary. In particular, the IAM amplitudes are built by
demanding i) exact unitarity and ii) that at low energies they match the
ChPT series to a given order. These conditions lead to the IAM
result, which is formally justified by the use of dispersion
relations \cite{unitvaciam}.  There is however a further, more
technical requirement, which is that the IAM partial-wave aplitudes should
vanish at the same values of the energy and with the same power as the
perturbative amplitudes. These values are the so called Adler zeros
and lie below threshold. Since a zero of the amplitude is a pole of
its inverse, this affects the analytic structure of $1/t$. A detailed
discussion  can be found in \cite{GomezNicola:2007qj} where the
proper correction to the IAM is derived using dispersion relations
and it is shown that these additional terms produce a negligible
effect in the physical region. However, as discussed in
\cite{FernandezFraile:2007fv}, taking into account this correction
is important when dealing with medium effects that can drive the
poles to the real axis, as it is the case here, since otherwise
there would be spurious poles both in the first and second Riemann
sheets below threshold.

The IAM can be extended at finite temperature by including the
thermal corrections to the scattering amplitude, which have been
calculated in  \cite{GomezNicola:2002tn} to one loop in ChPT. Since
temperature enters only in the loops, $t_2$ is $T$-independent. For
$t_4(s;T)$ one gets a perturbative unitarity relation exactly like
(\ref{pertunit}) but with the phase space replaced by:

\be \sigma_T (s)=\sigma_0(s)[1+2n_B(\sqrt{s}/2)]\label{thps}\ee with
$n_B(x)=[\exp(x/T)-1]^{-1}$  the Bose-Einstein distribution
function.

The function $\sigma_T(s)$ is the thermal phase space, which is
increased with respect to the $T=0$ one by the difference
$\left[1+n_B(E_1)\right]\left[1+n_B(E_2)\right]-n_B(E_1)n_B(E_2)=1+n_B(E_1)+n_B(E_2)$,
where $E_{1,2}$ are the energies of the two colliding pions,
corresponding to the difference between enhancement due to the
increase of two-pion outgoing states and absorption due to
collisions of the incoming pions with the thermal bath ones. In
the center of mass frame, where partial waves are defined,
$E_1=E_2=\sqrt{s}/2$ and the thermal phase space reduces to
(\ref{thps}). One can then use the same  $T=0$ IAM requirements,
replacing $\sigma_0\rightarrow\sigma_T$ and the partial waves by
the finite-$T$ ones, provided that only intermediate two-pion
states are relevant in the thermal bath, as expected in a dilute
gas regime at low and moderate temperatures. Finally, one arrives
to the thermal IAM formula for a given partial wave:

\ba t^{IAM}&=&\frac{t_2(s)^2}{t_2(s)-t_4(s;T)+A(s;T)}\nonumber\\
A(s;T)&=& t_4(s_2;T)-\frac{(s_2-s_A)(s-s_2)}{s-s_A}
    \left[t'_2(s_2)-t'_4(s_2;T)\right]\nonumber\\&&\ea
where the $A$ function is the Adler zero contribution discussed
above, $s_A$ denoting the Adler zero ($T$-dependent) expanded as
$s_A=s_2+s_4+\dots$ with $s_4=-t_4(s_2;T)/t'_2(s_2)$.

\begin{figure*}
\resizebox{0.5\textwidth}{!} {\includegraphics{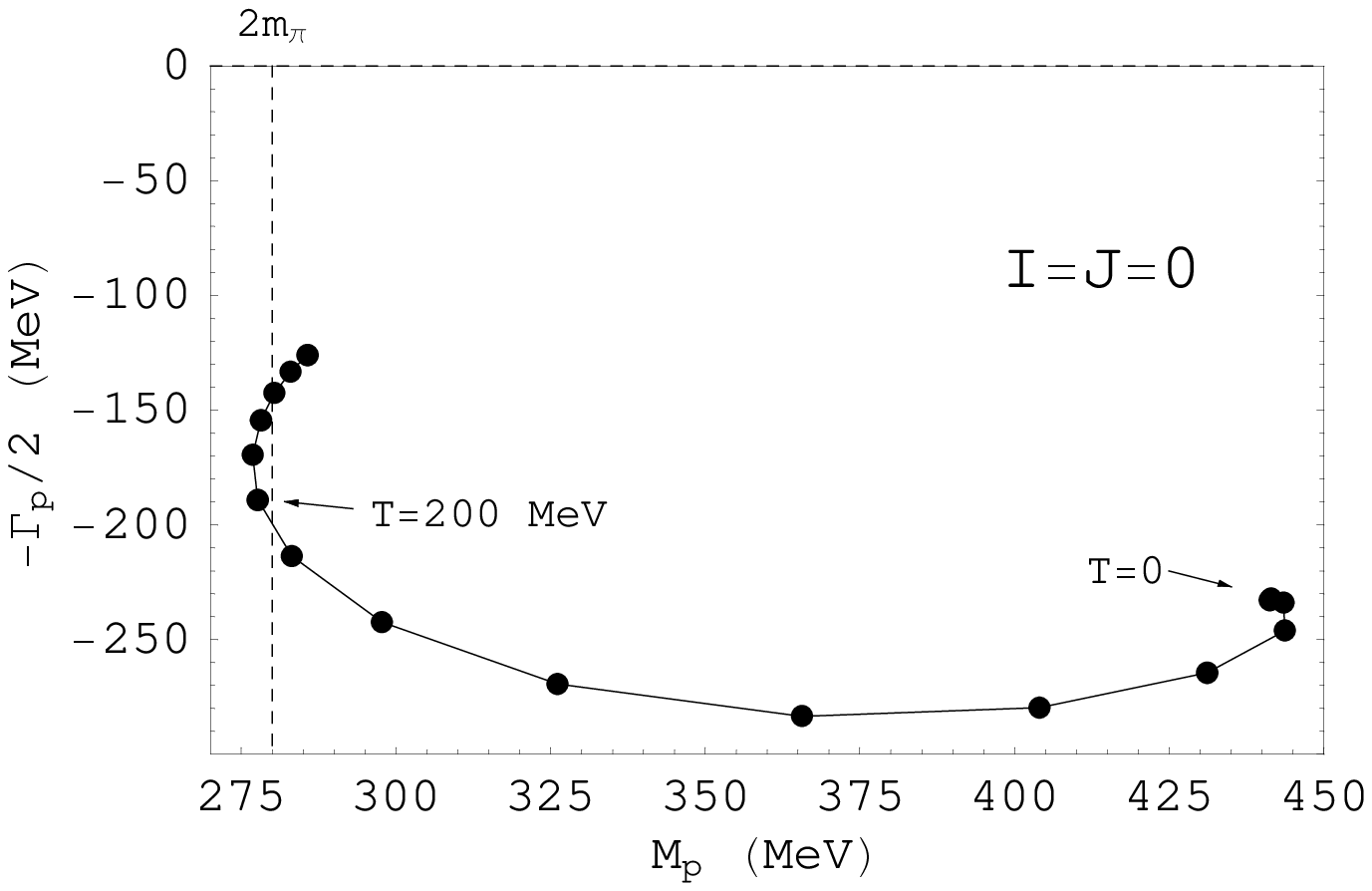}}
\resizebox{0.5\textwidth}{!} {\includegraphics{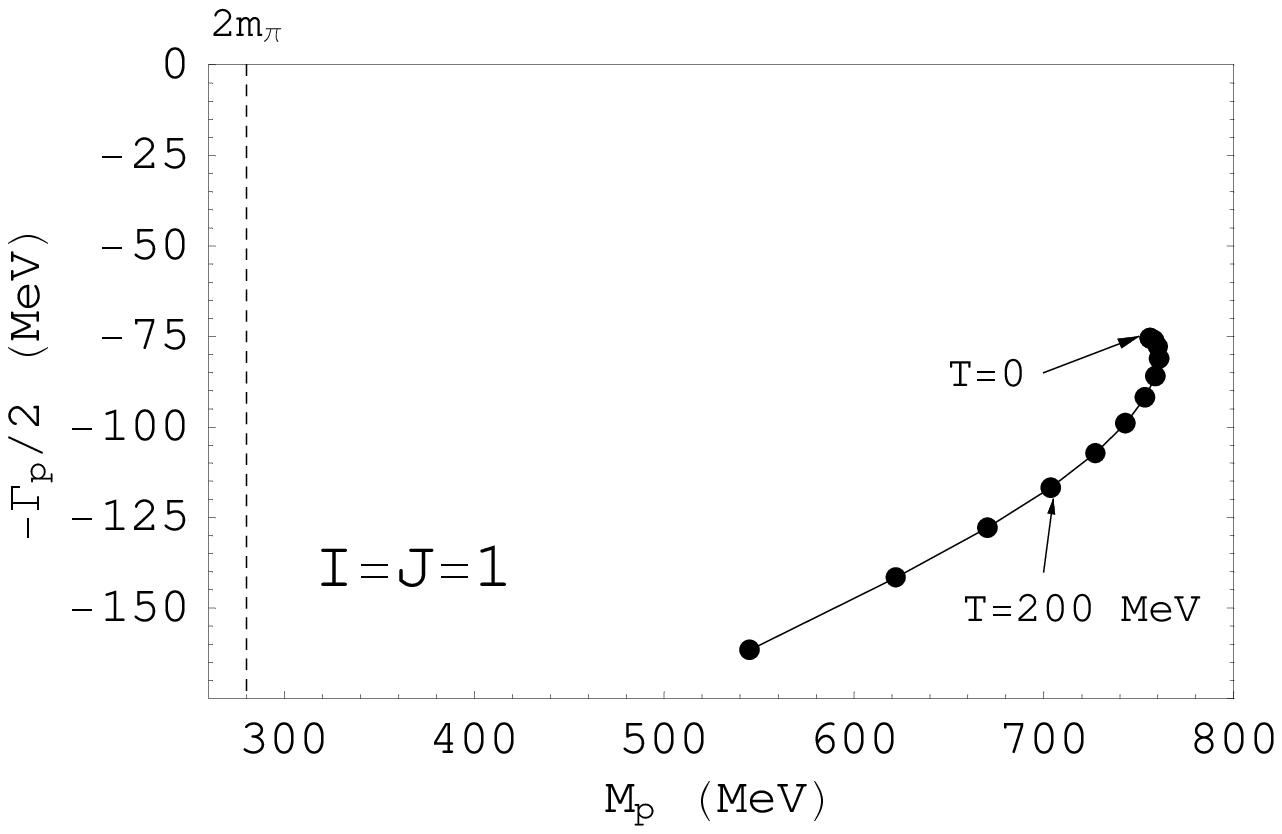}}% Here is how to import EPS art
%\vspace{-.3cm}
%\vspace*{5cm}       % Give the correct figure height in cm
 \caption{Temperature dependence of the  $f_0(600)$ and $\rho$
 complex poles unitarized by the IAM. The  points are obtained by varying the
temperature in 20 MeV intervals.} \label{fig:poles}
\end{figure*}

Performing the conventional extension of the amplitude to the second
Riemann sheet, one finds poles in the $I=J=0$ and $I=J=1$ channels
which are identified as the $f_0(600)$ and the $\rho (770)$. We show
in Figure \ref{fig:poles} the results for the pole position
$s_{pole}=(M_{p}-i\Gamma_p/2)^2$ for different temperatures. The
$\bar l_i$ values we have used are $\bar l_1=-0.3$, $\bar l_2=5.6$,
$\bar l_3=3.4$ and $\bar l_4=4.3$, which give for  the mass and
width of the $\rho(770)$ at $T=0$  $M_p\simeq 756$ MeV and
$\Gamma_p\simeq 151$ MeV. For the $f_0(600)/\sigma$ at $T=0$ we find
$M_p\simeq 441$ MeV and $\Gamma_p\simeq 464$ MeV.

The general features we observe are that the thermal $\rho$ pole
shows a predominant and increasing broadening behavior, while for
the $\sigma$ an important mass decrease takes place, presumably due
to chiral restoration, while the width increases for low
temperatures but decreases for temperatures of $T \simeq 100$~MeV and beyond.
In the rest of this section, we will discuss in more detail these
different behaviors in connection with the phenomenological issues
commented in the introduction.

\subsection{The thermal $\rho$ meson: broadening versus mass scaling in $\pi\pi$ scattering and dilepton probes.}
\label{sec:rhofinitetemp}

The $\rho$ pole obtained in our IAM thermal approach undergoes a
significant broadening at finite temperature. The main source of
thermal broadening is the Bose-Einstein increase of phase space
given in (\ref{thps}). However, it is not the only one. In fact,
using the Breit-Wigner ($\Gamma_p\ll M_p$) parametrization for the
$\rho$ exchange in $\pi\pi\rightarrow\rho\rightarrow \pi\pi$, one
gets \cite{Dobado:2002xf}:

\be
\frac{\Gamma_T}{\Gamma_0}=[1+2n_B(M_T/2)]\frac{g_T^2M_T}{g_0^2M_0}
\ee where $g$ is the effective $\rho\pi\pi$ vertex.

\begin{figure*}
\resizebox{0.52\textwidth}{!} {\includegraphics{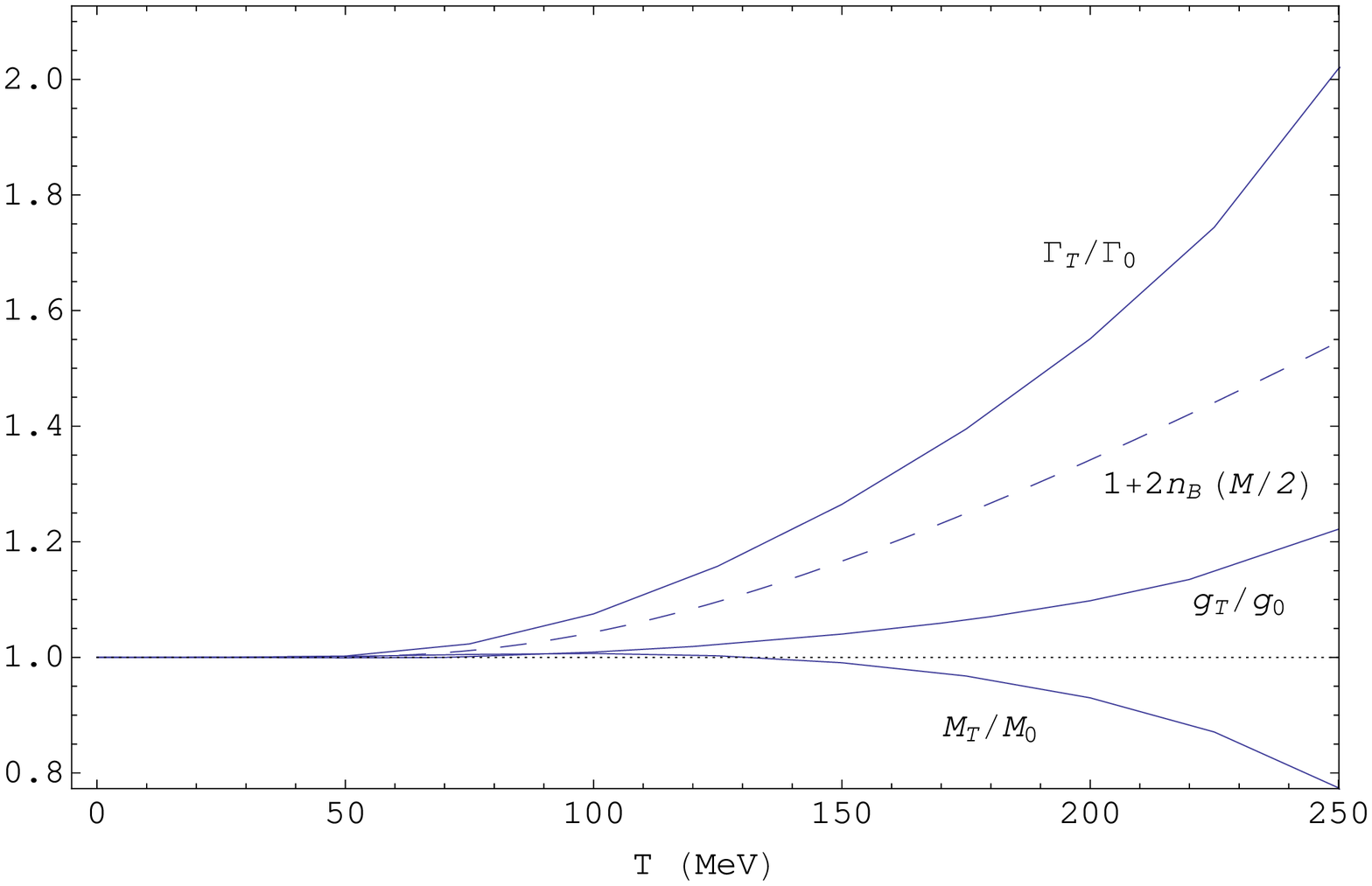}}
\resizebox{0.48\textwidth}{!} {\includegraphics{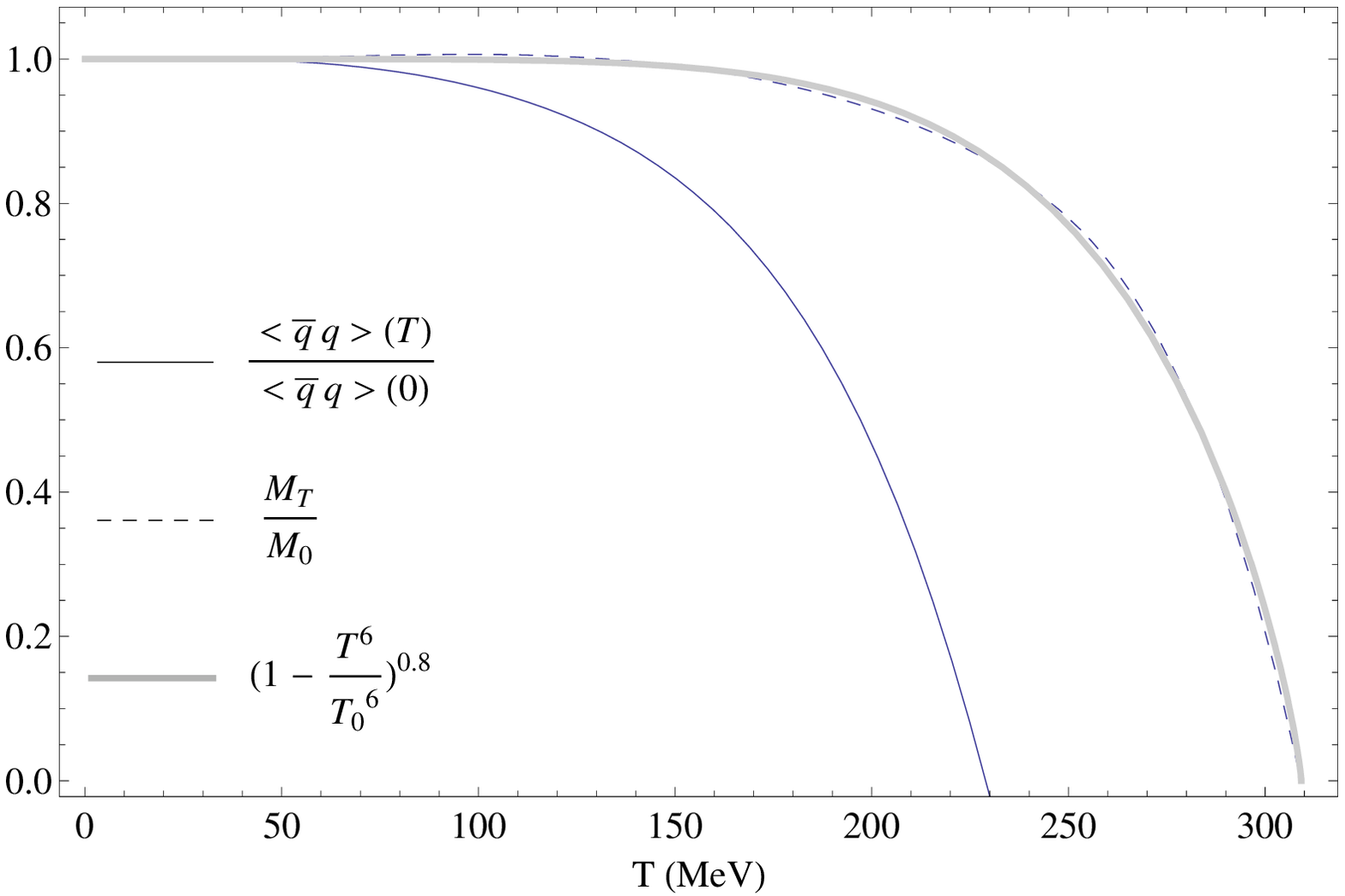}}% Here is how to import EPS art
%\vspace{-.3cm}
%\vspace*{5cm}       % Give the correct figure height in cm
 \caption{Thermal width and mass of the $\rho$ as extracted from its IAM pole. In the left panel we also show the
 $T$-dependence of the effective $\rho\pi\pi$ vertex. In the right panel we compare the mass with the quark
 condensate. The  mass vanishes at $T_0\simeq$ 310 MeV.} \label{fig:rhodetails}
\end{figure*}

In Figure \ref{fig:rhodetails} (left) we show in detail the
dependence of the mass, width and effective vertex and is clearly
seen that the width increases roughly as the phase space up to
$T\simeq$ 100 MeV and then further broadening arises due to the
increasing effective $\rho\pi\pi$ vertex. The  broadening we have obtained is
also present in the electromagnetic pion form factor \cite{glp05},
which enters directly into the dilepton yield (for back to back
dileptons in our case) arising from
$\pi^+\pi^-\rightarrow l^+l^-$. Our result is therefore
compatible with the broadening scenario observed clearly in NA60 \cite{NA60}
(see the discussion in the introduction).

The mass of the pole barely changes with temperature in our
approach. This seems to be in contradiction with the dropping-mass
scenario obtained in scaling models, which predict that the mass
should scale roughly with the quark condensate $M_T/M_0\simeq
\langle \bar q q \rangle_T/\langle \bar q q \rangle_0$
\cite{brownrho91}.
 However, it is worth mentioning that more recent analyses based on
 the same scaling hypothesis \cite{harada,br05} suggest that the mass dropping might be
 really effective only very close to the transition temperature,
 which in practice would mean that those predictions are not
 strictly incompatible with the dilepton data.

In order to clarify our results on this matter, we have plotted in
Figure \ref{fig:rhodetails} (right) the pole mass and the quark
condensate calculated from a virial expansion \cite{dopel9902} using
the $\Od(p^4)$ pion scattering amplitudes, which gives  a critical
temperature $T_c\simeq 230$ MeV. The results are extrapolated up to
the temperature $T_0$ where the mass vanishes, although our
ChPT-based approach is not meant to be valid there. We see that the
mass drops rather abruptly, with a $T^6$ power, while, as said
before,  it remains almost constant for $T$ below the chiral
transition. We do not see a scaling pattern when compared to the
condensate and, besides, $T_0\simeq$ 310 MeV lies far from the
critical value where the condensate vanishes. On the other hand, in
BR-like scaling models, the effective vertex decreases \cite{harada}
and, as commented before, there is no significant broadening. In
conclusion, although  we obtain a dropping mass qualitatively
compatible with BR-like models, our dominant broadening effect, the
increase of the effective vertex and the departure from the
condensate  are in conflict (at finite temperature) with that
scenario.

\subsection{The thermal $f_0(600)/\sigma$ meson: threshold behavior and
$\bar q q$  nature.} \label{sec:therf0}

As mentioned in the introduction, the interest in the $\sigma$ pole
concerns mainly its role as a precursor of the phase transition. The
temperature dependence we obtain in Figure \ref{fig:poles} shows
that mass decreasing is a prominent feature in this channel, thus
signaling chiral restoration. In fact, this is the dominant effect
here over phase space increase for $T \simeq 100$~MeV and beyond.
However, and this is crucial as
far as observable effects are concerned,  the pole remains broad
even when its real part has reached the two-pion threshold. The
reason behind this is that for a broad state ($\Gamma_p\sim M_p$)
the usual result $\Gamma\propto \sigma (M^2)\theta(M^2-4m_\pi^2)$,
which forbids the pole to be at threshold with a nonzero width,
should be replaced by $\Gamma\sim \int
\sigma(s)\theta(s-4m_\pi^2)\rho(s)$ \cite{FernandezFraile:2007fv}
where $\rho(s)$ is the resonance spectral function, different in
general from the narrow $\delta$-function
$\rho(s)=2\pi\delta(s-M^2)$. Therefore, the width can remain sizable
near threshold. In fact, a simple model of thermal width based on
this observation turns out to reproduce quite well the obtained IAM
poles \cite{FernandezFraile:2007fv}. This behavior for the $\sigma$
pole implies no threshold enhancement at finite temperature as a
precursor of the transition, since the pole is still far from the
real axis. This is clearly observed when we plot the squared modulus
of the $I=J=0$ partial wave in Figure \ref{fig:mod00} (left).
Threshold enhancement for a typical narrow resonance would mean for the
amplitude $\im
t(s) \sim 2m_\pi\theta\left(s-4m_\pi^2\right)/\sqrt{s-4m_\pi^2}$ for
$s\lsim 4m_\pi^2$ \cite{FernandezFraile:2007fv}. As we shall see
below, the situation changes dramatically when finite density
effects are included.

\begin{figure*}
\resizebox{0.5\textwidth}{!} {\includegraphics{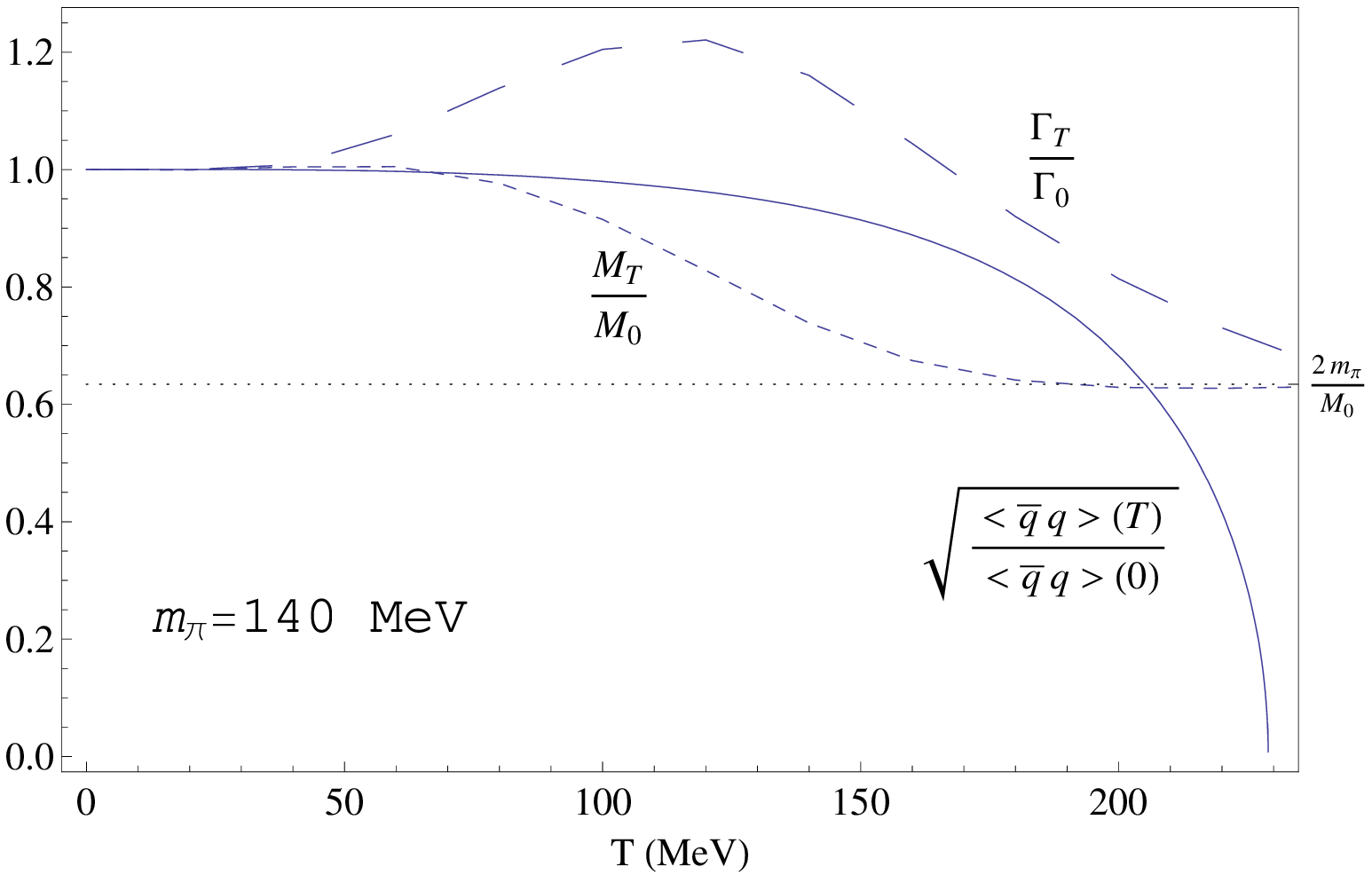}}
\resizebox{0.5\textwidth}{!} {\includegraphics{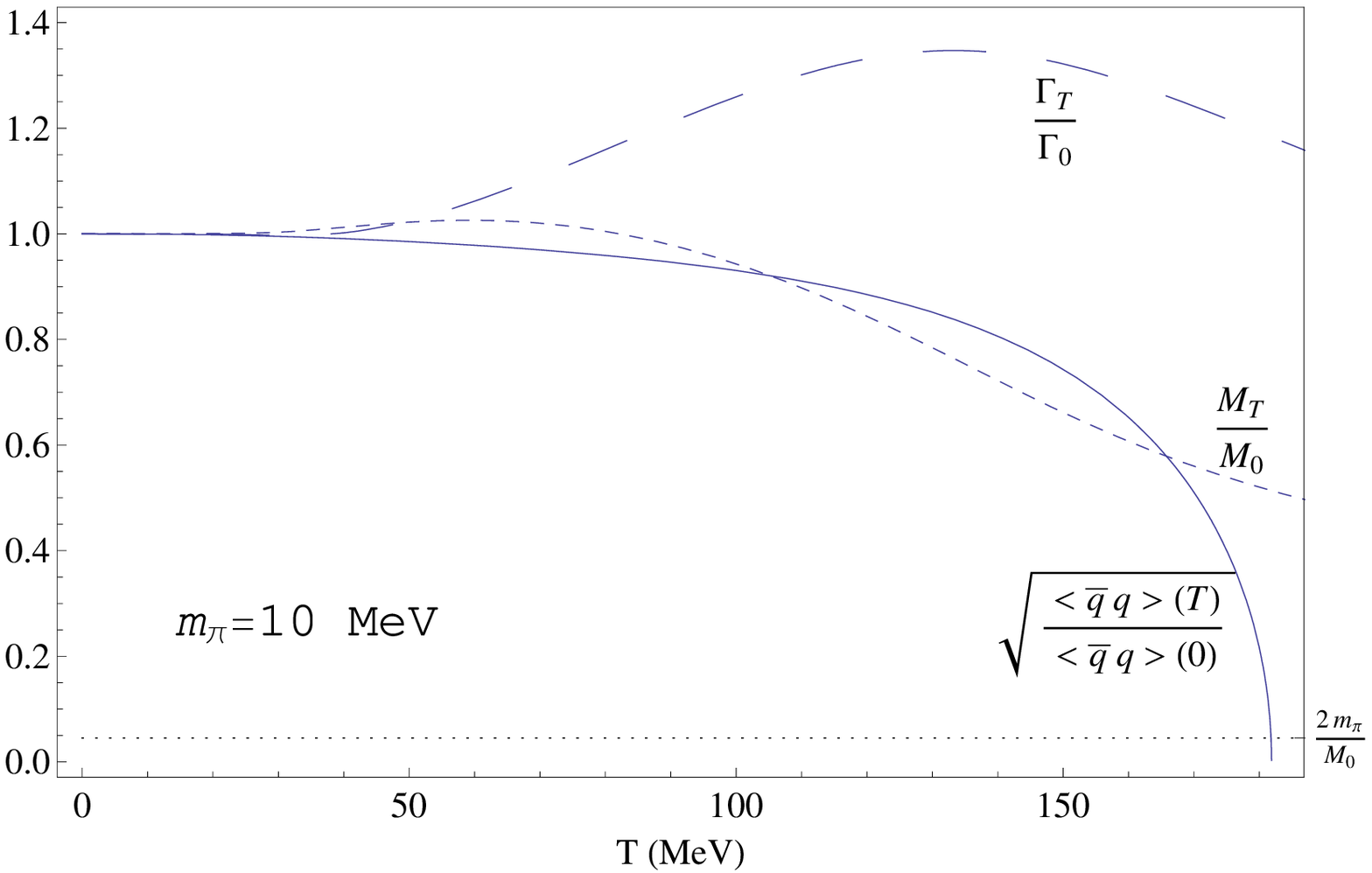}}% Here is how to import EPS art
%\vspace{-.3cm}
%\vspace*{5cm}       % Give the correct figure height in cm
 \caption{Comparison between  the $\sigma/f_0(600)$ mass (extracted from the IAM pole) at finite temperature and
  the root of the condensate, for the physical pion mass and near the chiral limit.} \label{fig:msigvscond}
\end{figure*}

The role of the $\sigma$ pole as a precursor of the transition is
strongly linked to its $\bar q q$ nature, as discussed in the
introduction. In Figure \ref{fig:msigvscond} we compare the mass
of the pole with the root of the quark condensate, similarly as we
did with the $\rho$ in the previous section. Recall that, in a
$O(4)$ model, $M_\sigma\sim \langle \sigma\rangle\sim f_\pi$ and,
on the other hand, $f_\pi^2=-m_q\langle \bar q q\rangle/m_\pi^2$
from the Gell-Mann-Oakes Renner (GOR) relation  \cite{gor}. We
also plot the results near the chiral limit, where the condensate
vanishes at a lower temperature and explicit chiral symmetry
breaking effects are minimized. In any case, we do not see a
scaling pattern, as expected from our result of a broad thermal
state. Near the chiral limit, the pole mass does not even go to
threshold near the critical temperature, since the width is
notably increased due to the more available phase space.
From our analysis of thermal effects one can also
conclude that the non-$\bar q q$ component of the
$f_0(600)$ must be of crucial importance. As discussed in the
introduction, the same conclusion has been reached in vacuum studies
\cite{Alford:2000mm,Pelaez:2003dy}. The novelty here is the use of
thermal arguments. We will see in Section \ref{sec:finitedenTzero}
that finite density effects driving the resonance poles to the real axis
below threshold allow also to extract interesting conclusions about
their $\bar q q$ nature.

\section{Finite density chiral restoring effects at $T=0$ through $f_\pi$ scaling in the IAM.}
\label{sec:finitedenTzero}

The simplest way to incorporate nuclear density effects at $T=0$ in
the unitarized chiral approach is to encode them only in the
variation of the pion decay constant, to linear order in density, as
\cite{thowir95oller02}:

\be \frac{f_\pi^2(\rho)}{f_\pi^2(0)}\simeq \frac{\langle \bar q
q\rangle(\rho)}{\langle \bar q q\rangle(0)}\simeq \left(1-
\frac{\sigma_{\pi N}}{m_\pi^2 f_\pi^2(0)}\rho\right)\simeq \left(
1-0.35\frac{\rho}{\rho_0}\right) \label{fpidensity} \ee where $\rho$
is the nuclear density, $\sigma_{\pi N}\simeq$ 45 MeV is the
pion-nucleon sigma term and $\rho_0\simeq$ 0.17 fm$^{-3}$ is the
normal or saturation nuclear matter density.

Therefore, by varying $f_\pi$ in our IAM approach, we mimic chiral restoring
nuclear effects \cite{FernandezFraile:2007fv}. This approach clearly ignores
standard many-body corrections, like the coupling of pions to particle-hole
($ph$) and Delta-hole ($\Delta h$) excitations, considered for instance in
\cite{Rapp:1997fs,Cabrera:2000dx,Chiang:1997di,cab05}. This approximation is
meant to be more adequate for the $\sigma$ than for the $\rho$ since, as we have
seen at finite temperature, chiral restoration  tends to dominate the $\sigma$
pole behavior under medium effects. In fact, by changing only $f_\pi$, no medium
broadening is produced, which is probably unrealistic for the $\rho$ case, as
emphasized in many-body works
\cite{Herrmann:1992kn,Peters:1997va,Cabrera:2000dx}. This must be borne in mind
when interpreting our results for the $\rho$ in terms of BR-scaling. In any
case, in Section \ref{sec:finitedentemp} we will consider a different
unitarization scheme, which allows to introduce all the above mentioned nuclear
many-body effects and compare with our simple ``$f_\pi$ scaling'' considered
here.

\begin{figure*}
\resizebox{0.46\textwidth}{!} {\includegraphics{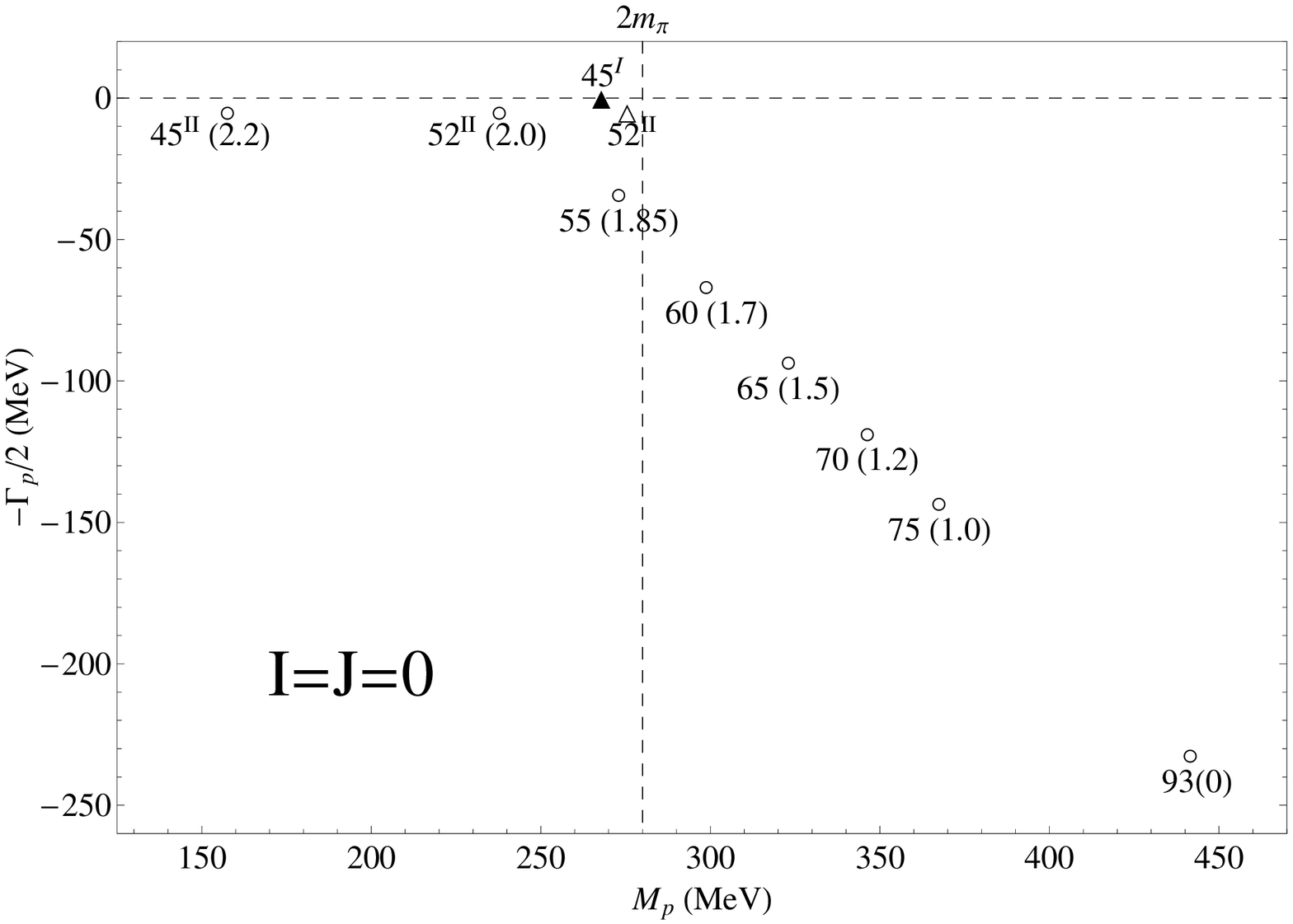}}
\resizebox{0.54\textwidth}{!} {\includegraphics{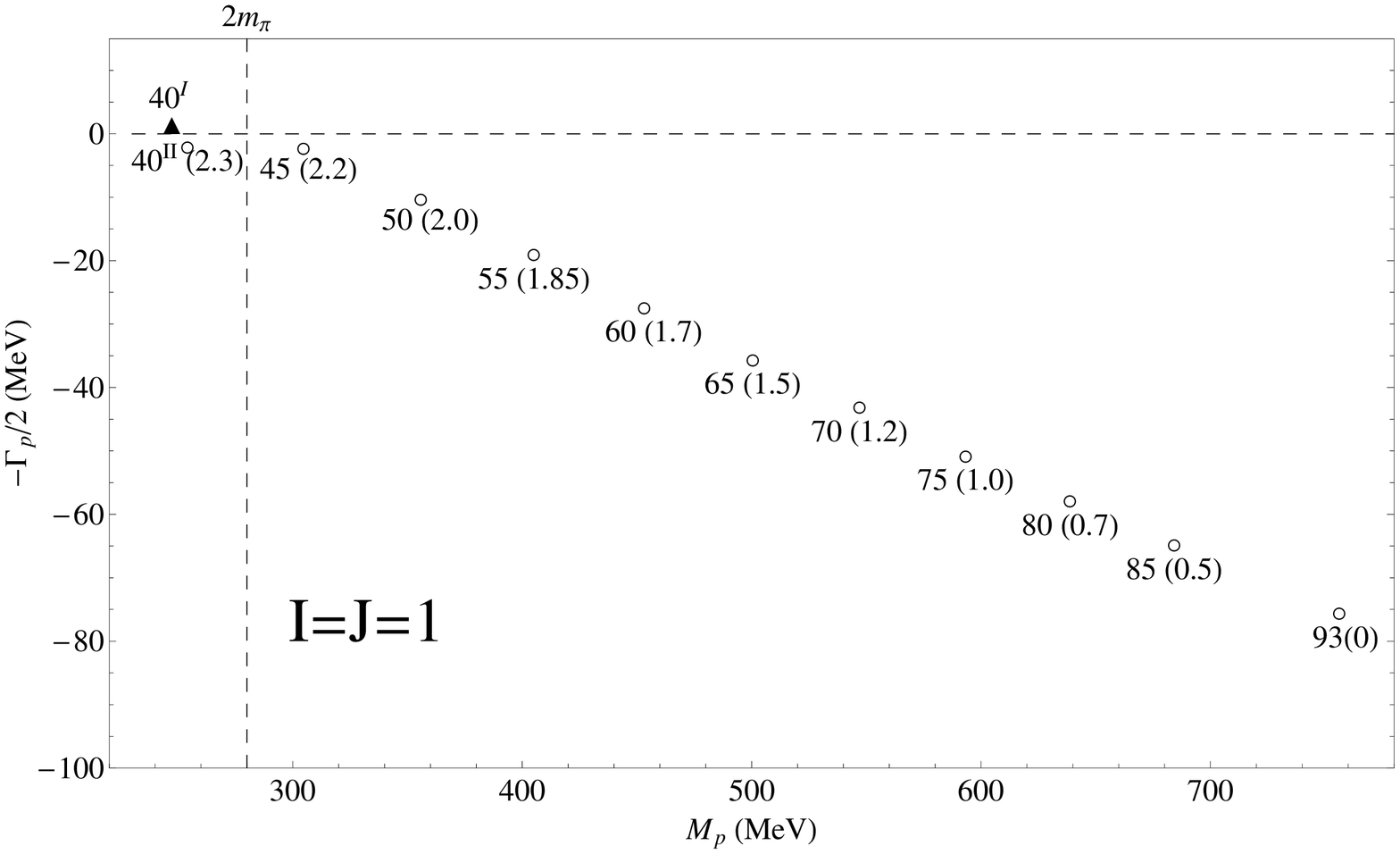}}
% Here is how to import EPS art
%\vspace{-.3cm}
%\vspace*{5cm}       % Give the correct figure height in cm
 \caption{IAM poles with varying $f_\pi$. The numbers attached to the pole positions indicate
 $f_\pi (\rho)(\rho/\rho_0$ in brackets) in MeV according to (\ref{fpidensity}). The circles and open triangles denote second sheet poles,
 while black triangles refer to first sheet ones.} \label{fig:polesdenfpi}
\end{figure*}

\begin{figure*}
\resizebox{0.5\textwidth}{!} {\includegraphics{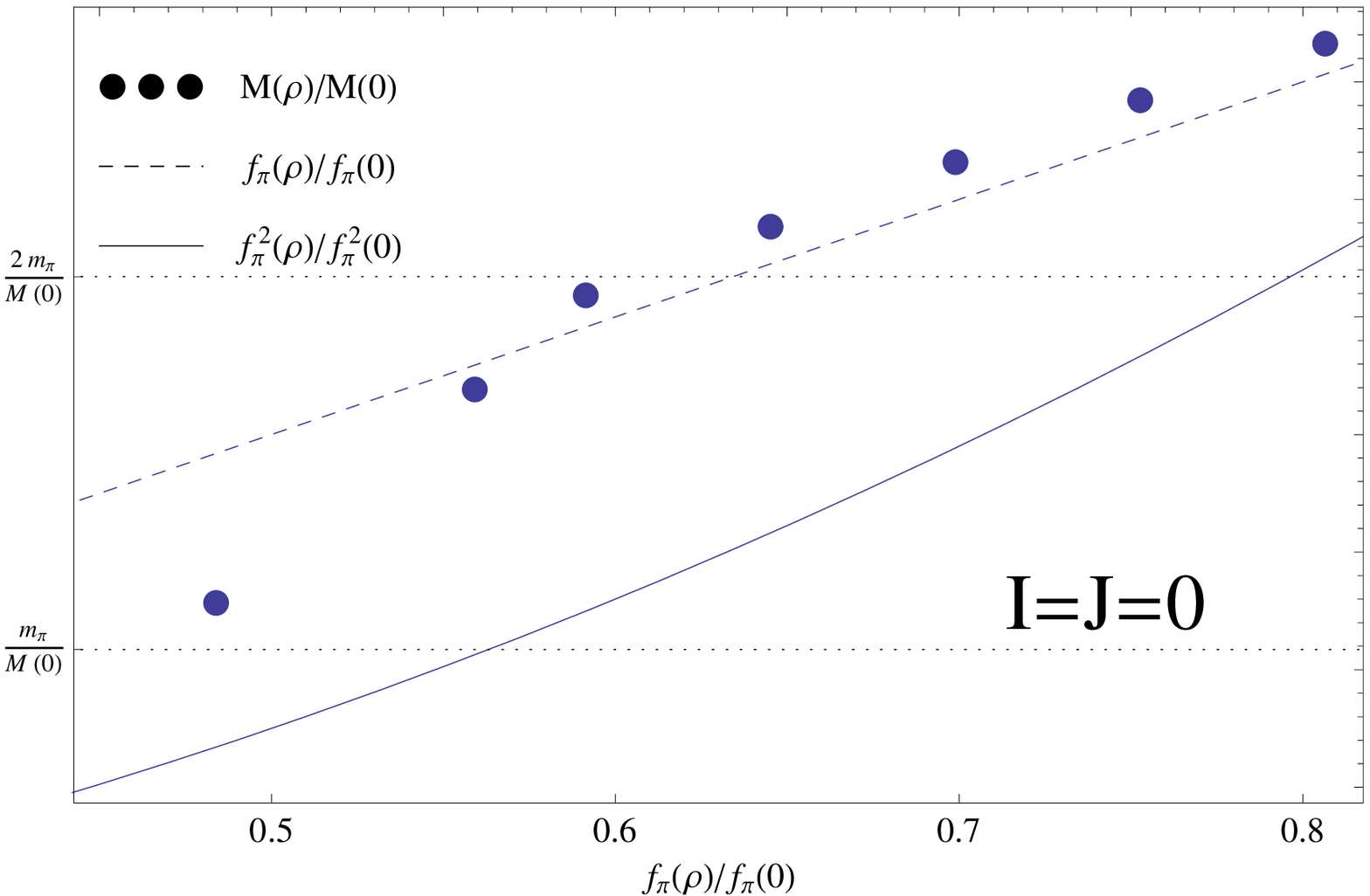}}
\resizebox{0.5\textwidth}{!} {\includegraphics{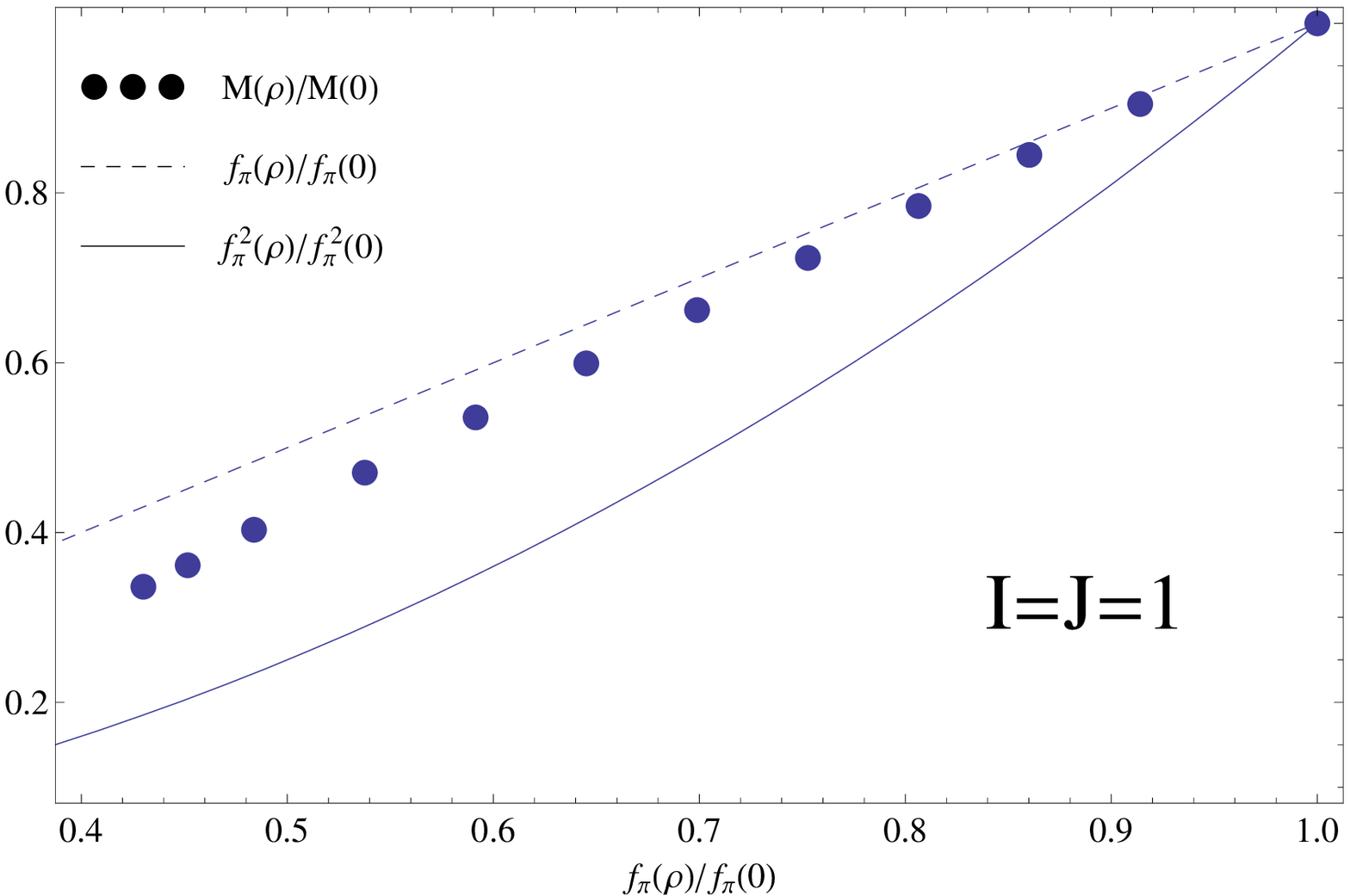}}
% Here is how to import EPS art
%\vspace{-.3cm}
%\vspace*{5cm}       % Give the correct figure height in cm
 \caption{Scaling of the mass, compared with  $f_\pi^2(\rho)/f_\pi^2(0)\sim \langle \bar q q\rangle(\rho)/\langle
 \bar q q\rangle(0)$
 and with $f_\pi(\rho)/f_\pi(0)$. In the $I=J=0$ case, the masses displayed correspond to the lowest masses of the
 second-sheet poles.} \label{fig:scalingfpi}
\end{figure*}

\begin{figure*}
\resizebox{0.5\textwidth}{!} {\includegraphics{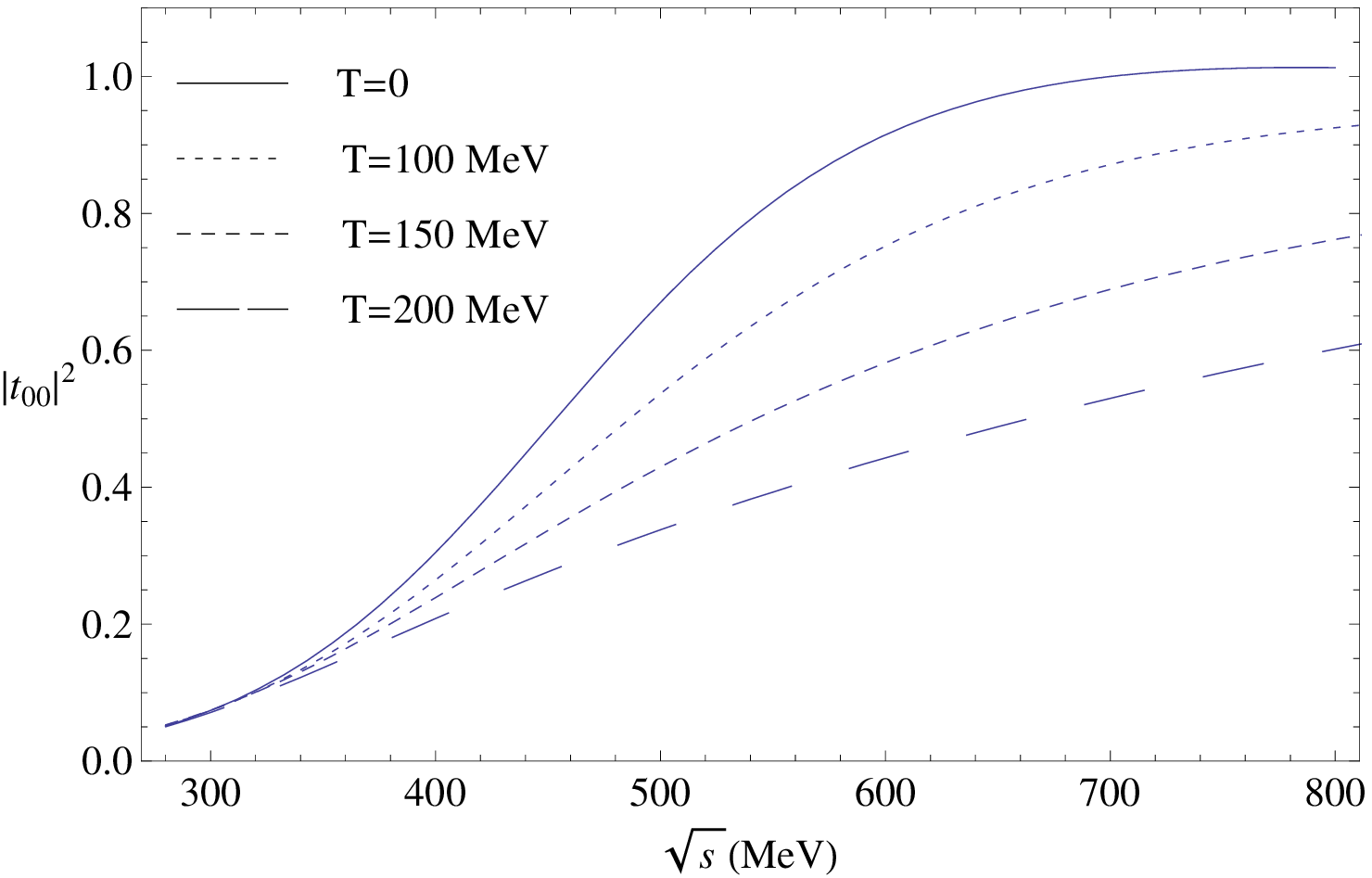}}
\resizebox{0.5\textwidth}{!} {\includegraphics{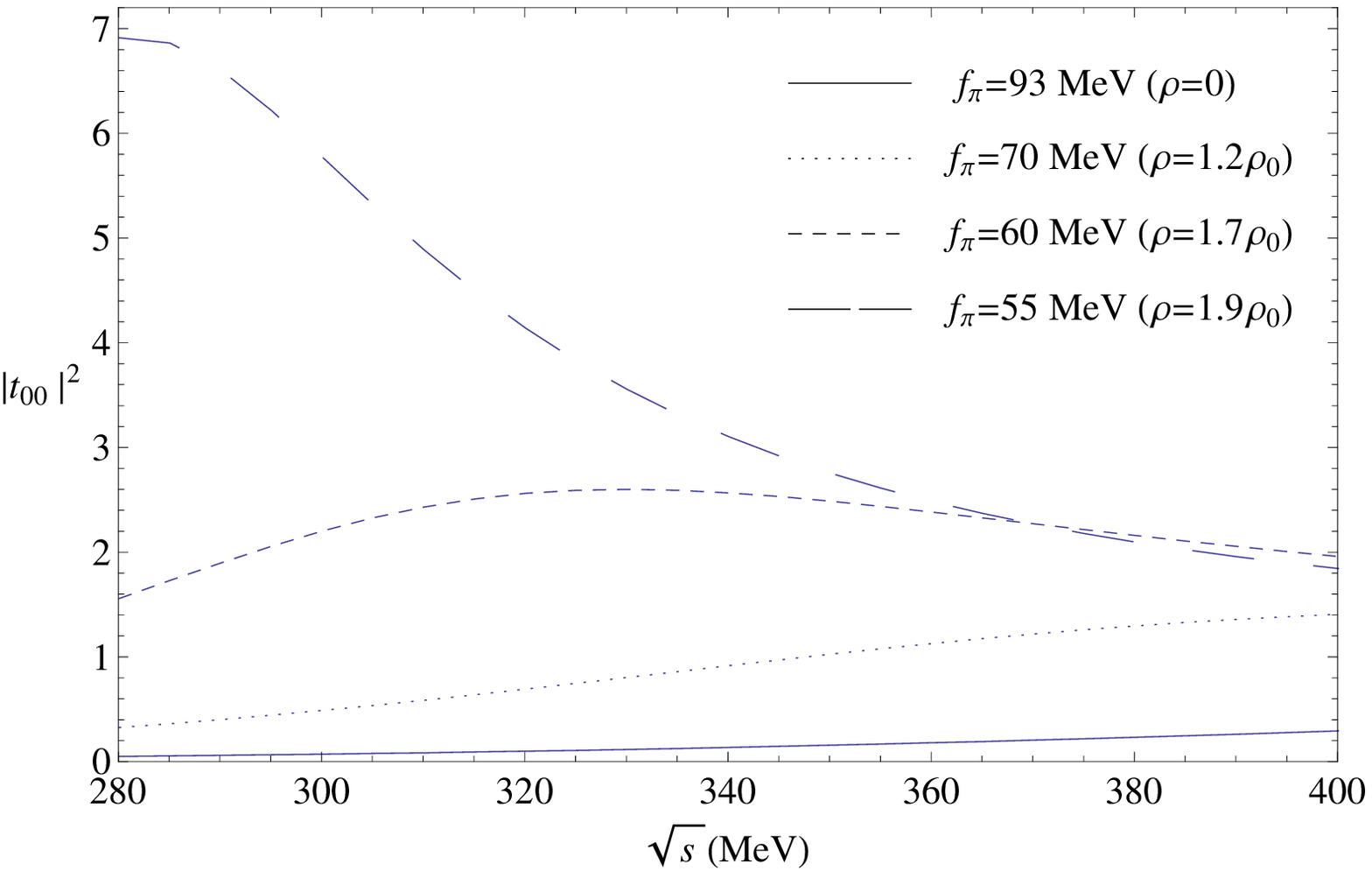}}% Here is how to import EPS art
%\vspace{-.3cm}
%\vspace*{5cm}       % Give the correct figure height in cm
 \caption{Squared modulus of the $I=J=0$ partial wave at finite temperature (left) and at $T=0$ and
  varying $f_\pi$ (right) to simulate  chiral restoration at finite density.} \label{fig:mod00}
\end{figure*}

\begin{figure*}
\resizebox{0.5\textwidth}{!} {\includegraphics{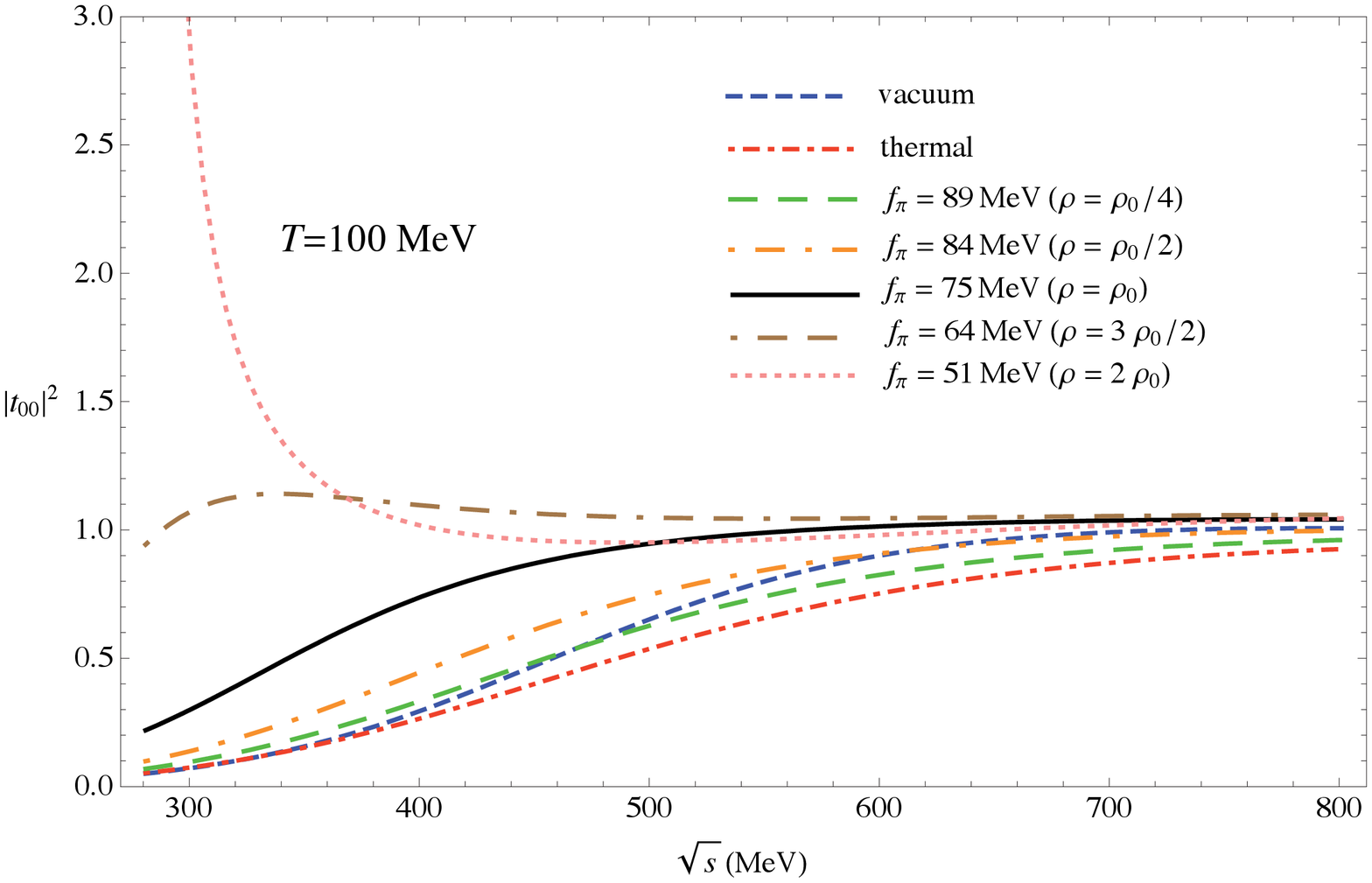}}\resizebox{0.5\textwidth}{!} {\includegraphics{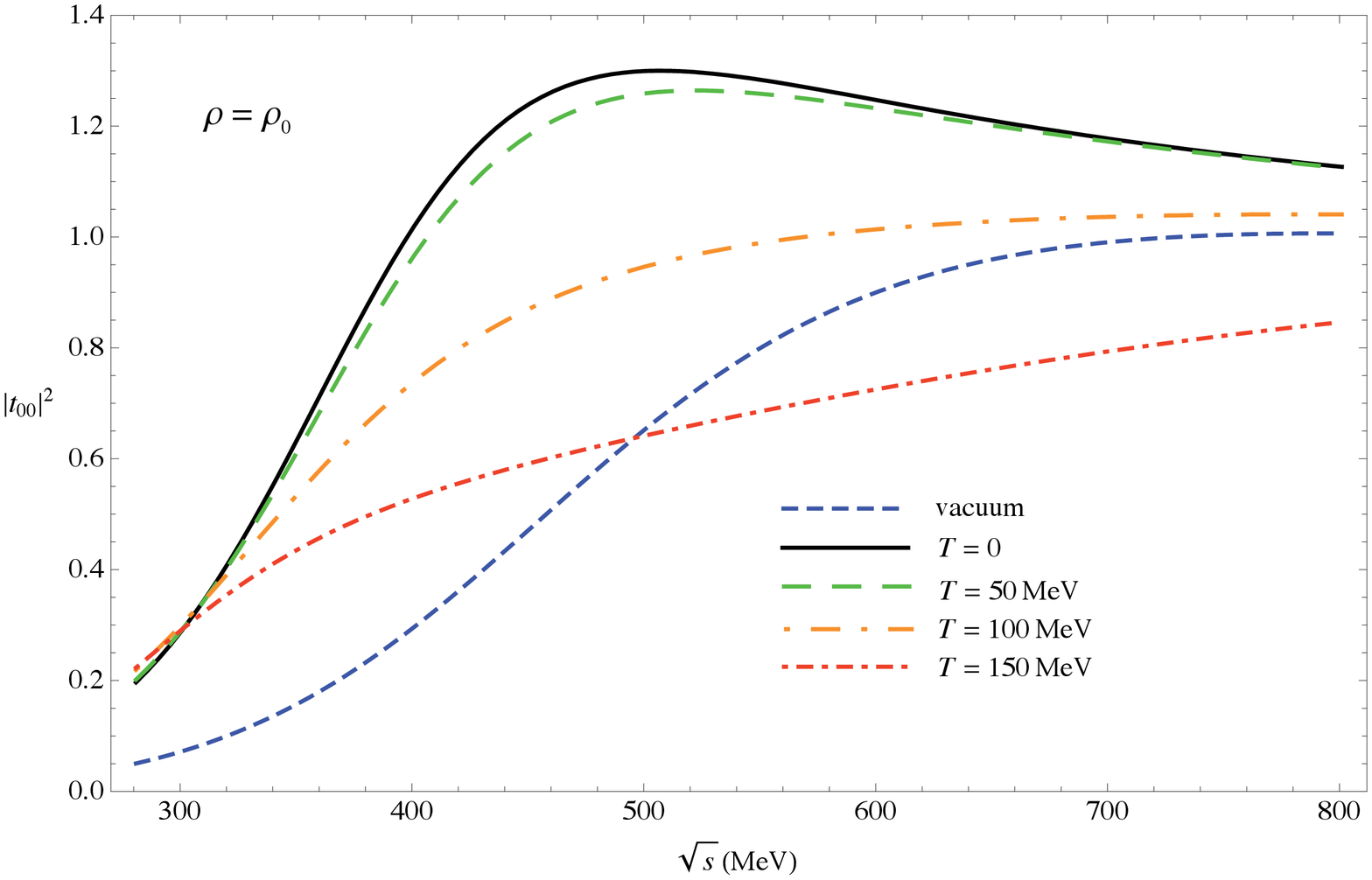}}
\caption{Squared modulus of the $I=J=0$ partial wave at $T=100\ \mathrm{MeV}$
for several densities (left), and at nuclear matter density ($f_\pi=75$~MeV)
for several temperatures (right).} \label{fig:finiteTFPI}
\end{figure*}

The results obtained by varying $f_\pi$ in the IAM amplitudes are
displayed in Figs.~\ref{fig:polesdenfpi} and \ref{fig:scalingfpi}
for the pole trajectories in the complex plane and mass scaling,
respectively, in the 00- and 11-channels. In Fig.~\ref{fig:mod00}
(right) we also show the effect on the scattering amplitude in the
$\sigma$ channel. As $f_\pi$ decreases the $\sigma$ pole becomes
narrow enough so that chiral mass reduction brings it to the real
axis, which produces threshold enhancement in the amplitude although
at densities well above $\rho_0$. Regarding the mass dropping, it
takes place now along  with the condensate (see
Fig.~\ref{fig:scalingfpi}). In Fig.~\ref{fig:finiteTFPI} we show
$|t_{00}|^2$ considering simultaneously temperature and finite
density according to Eq.~(\ref{fpidensity}) for $f_\pi$. We observe
that, for a given value of $f_\pi$ (small enough), the net effect of
introducing temperature is to amplify threshold enhancement, as the
amplitude is notably softened at higher energies, by thermal
broadening, what makes the low energy region relatively more
important. Still, one has to keep in mind that (a) this effect is
driven by the proximity of the $\sigma$ pole to the real axis when
decreasing $f_\pi$, (b) thermal effects alone do not generate
threshold enhancement in the amplitude as discussed above, and (c)
the $\pi\pi$ phase space is not open below $2m_\pi$ even at finite
temperature. In Sect.~\ref{sec:finitedentemp} we will compare these
results for the 00-channel with an implementation of nuclear density
effects in a dynamical many-body calculation.

As for the $\rho$
channel, the BR-like scaling pattern is now closely followed. In fact, our
$\rho$-meson pole moves gradually from the $\sqrt{\langle \bar q q\rangle}$
curve to the $\langle \bar q q\rangle$ one, as obtained in \cite{br04}.
However, one must be careful about this conclusion, since we are disregarding
medium related broadening which might change the scaling picture observed here.
Strictly speaking, our results indicate that if the relevant density effects
amount {\em only} to a scaling of  $f_\pi$, then one gets scaling in the pole
mass, which is quite consistent with the BR idea.

There is an aditional interesting feature of our results: when density is
increased further, $\pi\pi$ bound states (first sheet poles) appear
just below threshold in both channels. As we shall see in
Sect.~\ref{sec:molec}, this result allows for an
interpretation in terms of a ``molecular'' classification of
resonances, which is completely different in both channels. In fact,
we note that in the $\sigma$ channel the bound state is preceded by a
doubling of poles in the second sheet. This will be further discussed in
Sect.~\ref{sec:molec}.
The appearance of pole
doubling and bound states has been also analyzed in other works
\cite{patkos03,hidaka04,Schuck:1988jn}.

\subsection{Comparison with nuclear matter experiments.}
\label{sec:compexp}

As discussed above, apart from Heavy Ion
Collisions, there are several dedicated experiments on production reactions in
nuclei where our analysis can be useful, namely, those related to threshold
enhancement in $\pi\pi$ production
and the modification of dilepton spectra from $\rho$ decays.

The threshold enhancement observed in
\cite{Bonutti:2000bv,cb,messetal} in the $I=J=0$ channel is
qualitatively reproduced in our ``$f_{\pi}$ scaling'' finite-density approach.
Comparing with our finite-$T$ analysis in the previous sections, our
main conclusion is that, due to the broad nature of the $\sigma$ at
finite $T$, threshold enhancement is not visible at finite $T$ and
it can only take place if chiral restoring finite-density effects
are strong enough, compared for instance with medium effects  we
have neglected  which could also increase the $\sigma$ width,
preventing the pole from approaching the real axis, as in the
finite-$T$ case, and erasing the threshold effect. This does not
seem to be the case when many-body interactions are properly
accounted for \cite{Chiang:1997di,roca02}, which gives support to this simple
approximation. In fact,  the observed effect is in reasonable
quantitative agreement with previous theoretical works
\cite{davesne00,jihatku01,patkos03} and it is not very strong for
densities $\rho\lsim \rho_0$ (the range available experimentally) in
line with the experimental results in \cite{cb}.

As for dileptons in nuclear matter, a parameter directly measurable
 in those experiments is the coefficient of the linear density term for the  mass of the $\rho$(770) meson:

 \be
\frac{M(\rho)}{M(0)}=1-\alpha\frac{\rho}{\rho_0}
 \ee

The experimental values available so far are $\alpha=0.092\pm 0.002$
obtained by the E325-KEK collaboration \cite{Naruki:2005kd} and
$\alpha=0.02\pm 0.02$ measured by the JLab-CLAS experiment
\cite{:2007mga}. We have performed a linear fit of our pole results
in the $I=J=1$ channel for $f_\pi=93,85,80,75$~MeV, i.e, up to
$\rho\lsim\rho_0$ to be closest to the experimental situation. Our
fit gives $\alpha=0.2$, which is closest to the result in
\cite{Naruki:2005kd} than to  \cite{:2007mga}, although a bit above
the experimental value. Our result is also in agreement with
Brown-Rho scaling \cite{br02} and QCD sum rules
\cite{Hatsuda:1995dy} which predict $\alpha\simeq 0.1-0.2$. It is
important to remark again that so far we ignore all medium
broadening effects, which, unlike the case of the $\sigma$, may be
crucial in this case, as emphasized in many-body treatments
\cite{Herrmann:1992kn,Peters:1997va,Cabrera:2000dx,Urban:1998eg}. In
fact,
 in \cite{Leupold:1998bt} it was realized that QCD sum rules
 themselves do not provide a unique constraint for the in-medium mass and
 width variation, unless additional model assumptions are made. In
 particular, it is shown that if one assumes that the
 width is not increased then automatically the mass drops. This is
 the scenario we have recovered in our present approach, which, as
 commented several times above, does not mean that this is the
 physically relevant case.

\subsection{``Molecular'' classification of resonances.}
\label{sec:molec}

One of the objectives of this work is to analyze the $\bar q q$
structure of light meson resonances from a thermal and finite-density
viewpoint. We have already discussed in Section  \ref{sec:therf0}
that our results for the thermal $\sigma$ are not consistent with
its pure $\bar q q$ nature.

The situation changes qualitatively with the $T=0$ finite-density
dependence obtained in Section \ref{sec:finitedenTzero}. We see in
Figure \ref{fig:scalingfpi} that the $\sigma$ pole follows quite
well the same pattern of a $O(4)$ $\sigma$-like $\bar q q$ state.
One may then wonder about the implication of this for the $\bar q q$
nature of the in-medium $\sigma$.  We want to point out  here that
actually one can gain very useful information about that by looking
at the behavior of the poles near threshold. Our argument is
supported by the classification of resonances lying near threshold
given in \cite{Morgan:1990ct} and based on the effective range
approximation. In those works it is stated that although generally
it is difficult to extract properties about the ``internal'' nature
of resonances from its decay products (scattering poles), when the
pole lies near threshold, a general rule can be applied: a
``potential'' or ``molecular'' resonance shows up as a single pole
near threshold, while for a $\bar q q$-like state two poles near
threshold appear very close to another in different Riemann sheets.
This classification was originally applied to states like the $f_0
(980)$, which lies very close to the $\bar K K$ threshold.

Our claim here is that this classification argument can equally be
applied to light resonances if medium effects drive the poles to the
real axis. This is exactly the situation when we include density
effects only through Eq.~(\ref{fpidensity}), as it is clear from the
results in Figure \ref{fig:polesdenfpi}, although bearing in mind
that the effective values of $f_\pi$ for which the poles reaches the
real axis are too small to trust entirely our ChPT-based approach.
We see that the $\sigma$ pole follows a clear ``molecular'' pattern,
since the pole that remains close to threshold and eventually
becomes a bound state is well separated from the second-sheet pole
that lies below threshold. We interpret this as a coexistence
of two states for high enough densities: a $\pi\pi$ ``molecule'' and
a virtual state which behaves as a ``chiral partner'' of the pion, in
the sense that it tends to become degenerate in mass with it (albeit
with different quantum numbers) following the order parameter. Note
that the pole we represent in Figure \ref{fig:scalingfpi} is
precisely this virtual state. This picture is in contrast with what
we observe for the $\rho$ channel, where two nearby poles remain
below threshold, one of them moving to the first sheet and becoming
a bound state. This is clearly a $\bar q q$, non-molecular scenario,
according to the previously discussed classification. We remark that
a similar picture for the real axis poles has been obtained in
\cite{patkos03}, where the density dependence is also parametrized
in $f_\pi$ and in vacuum by increasing the quark mass in order to
compare with lattice results \cite{Hanhart:2008mx}.

\section{Finite temperature and density in a many-body unitarized approach.}
\label{sec:finitedentemp}

In this Section we extend our previous results in the $I=J=0$ channel by
incorporating finite nuclear density and temperature
effects in a many-body description of pion dynamics.
We follow the line of Refs.~\cite{Chiang:1997di,cab05} where the effective
$\pi\pi$
scattering amplitude in cold nuclear matter was evaluated in a chiral unitary
framework. For technical reasons which we discuss below, it is convenient to
use a different unitarization scheme for the $t$-matrix, which however provides
similar results in vacuum to those from the IAM, i.e., the light
meson-meson resonances are dynamically generated in the scalar channel and many
scattering observables are described in good agreement with
experiment \cite{Oller:1997ti,Oller:1998hw,Oller:1997ng}.

\begin{figure}
\begin{center}
\resizebox{0.4\textwidth}{!} {\includegraphics{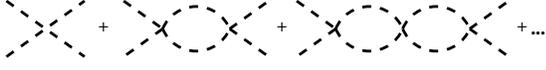}}
% Here is how to import EPS art
%\vspace{-.3cm}
%\vspace*{5cm}       % Give the correct figure height in cm
 \caption{Diagrammatic representation of the Bethe-Salpeter equation for
 $\pi\pi$ scattering.}
 \label{fig:diagrams}
 \end{center}
\end{figure}

The idea is to solve the coupled-channel Bethe-Salpeter (BS) equation for the
partial wave scattering amplitude (in matrix notation),
\begin{equation}
\label{eq:LS}
T =  V + \overline{V G T} \, ,
\end{equation}
where the potential (kernel) $V$ of the equation
is provided by the lowest order tree level amplitude from the chiral Lagrangian.
In Eq.~(\ref{eq:LS}),
\begin{eqnarray}
\label{G_vacuum}
G_i (P) &=& {\rm i} \,
\int \frac{d^4q}{(2\, \pi)^4} \,
\frac{1}{(P^0-q^0)^2 - \vec{q}\,^2 - m_i^2 + {\rm i} \varepsilon}
\nonumber \\
&\times&
\frac{1}{(q^0)^2 - \vec{q}\,^2 - m_i^2 + {\rm i} \varepsilon}
\end{eqnarray}
stands for the intermediate two-particle meson-meson Green's
function of channel $i$ ($G$ is diagonal), where $P=(P^0,\vec{0})$ is the total
external momentum in the center of mass frame of the two pions (rest frame with
respect to the nuclear medium), with $s=(P^0)^2$.
In principle, both $V$
and $T$ enter off-shell under the momentum integration
($\overline{VGT}$ term) of the meson-meson loop. However, as it
was shown in Ref.~\cite{Oller:1997ti}, the (divergent)
off-shell contributions of $V$ and $T$ in the $s$-wave interaction
can be reabsorbed in a renormalization of the bare coupling
constants and masses order by order. Therefore, both $V$ and $T$
can be factorized on-shell, leaving
the four-momentum integration only in the two-particle
meson-meson propagator, cf.~Eq.~(\ref{G_vacuum}).
An alternative justification for solving
the Bethe-Salpeter equation with on-shell amplitudes can be found
in the framework of the $N/D$ method, applied for meson-meson
\cite{Oller:1998zr} and meson-baryon
\cite{Oller:2000fj} interactions. We are thus left with a
set of linear algebraic equations with trivial solution,
\begin{equation}
T = [1 - V G ]^{-1} V \,\,\, .
\label{eq:BSalgeb}
\end{equation}
Since we are interested in the $\sigma$ meson energy region (and particularly close
to the two-pion threshold), for simplicity we shall work in a single-channel
approach (the coupled $\bar K K$ channel lies far above in energy and has little
effect at low energies). The conditions under which this unitarization
procedure is
equivalent to the IAM were discussed in \cite{Oller:1998hw,Oller:1997ng}.
The main
difference of the present approach as compared to the full IAM amplitude lies in
the ${\cal O}(p^4)$ contribution (which here comes from the $s$-channel
meson-meson loop and
no tadpole or ${\cal L}_4$ tree-level terms are included) and the absence
of $t$- and $u$-channel diagrams
(or, in other words, the left-hand analytical cut is ignored). Still, the
present scheme dynamically generates the $\sigma$ pole with similar properties
as in the IAM and the experimental phase shifts in $\pi\pi$ scattering
are well reproduced
\cite{Oller:1997ti,Oller:1998hw,Oller:1997ng}.
Also note that Eqs.~(\ref{unit},\ref{thps})
are equally satisfied in this approach.
We show in Fig.~\ref{fig:diagrams} a diagrammatic representation of the series
of $s$-channel diagrams which is summed in the BS equation.

\begin{figure}
\begin{center}
\resizebox{0.2\textwidth}{!} {\includegraphics{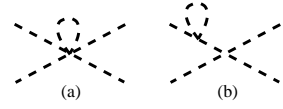}}
% Here is how to import EPS art
%\vspace{-.3cm}
%\vspace*{5cm}       % Give the correct figure height in cm
 \caption{Pion tadpole diagrams in $\pi\pi$ scattering.}
 \label{fig:tadpoles}
 \end{center}
\end{figure}

The calculation of the thermal amplitude proceeds by first
reanalyzing the on-shell factorization of $V$ and $T$ in the BS
equation. At
finite temperature, the divergent contribution from off-shell terms in the
one-loop amplitude has the
same structure as in vacuum, but a finite, temperature-dependent part survives
which cannot be cast as a (vacuum) renormalization of $f_{\pi}$ or $m_{\pi}$.
This contribution can be accounted for as a temperature correction
to the ${\cal O}(p^2)$ kernel,
\begin{eqnarray}
\label{deltaVoff}
\delta V_{\rm off}^T (s)= \frac{4}{3 f_{\pi}^2} \left( V_{\rm on}(s)
+ \frac{s}{3 f_{\pi}^2} \right) \, I_0^T \ ,
\end{eqnarray}
with $I_0^T = (2 \pi^2)^{-1} \int_0^{\infty} dq \, \vec{q}\,^2 \,
n_B (\omega_q) / \omega_q$, $V_{\rm on}(s) = -
(s-m_{\pi}^2/2)/f_{\pi}^2$ and $\omega_q^2 = \vec{q}\,^2 +
m_{\pi}^2$ (we follow in this section the normalization of partial
waves given in \cite{Oller:1997ti,Oller:1998hw}, which differs from
the one in Sect.~\ref{sec:finitetemp} in a factor $16\pi$). In order
to keep as close as possible to the physics described by the IAM
amplitude, we have also considered the finite-$T$ contribution from
tadpole terms, which becomes relevant as the $\sigma$ meson pole is
driven towards the two-pion threshold and its behavior is no longer
dominated by the two-pion phase space \cite{GomezNicola:2002tn}. The
corresponding diagrams are depicted in Fig.~\ref{fig:tadpoles} and,
to the lowest order in the chiral counting, they emerge from ${\cal
O}(p^2)$ interaction terms with up to six meson fields. The
finite-$T$ correction to the tree level amplitude to account for
these terms reads
\begin{eqnarray}
\label{deltaVtadwrf}
\delta V_{\rm tad}^T (s) = \left[ \frac{20}{9}\,s\,I_0^T - \frac{25}{6}\,I_2^T
\right] / f_{\pi}^4 + V_{\rm on} (s) \frac{4}{3 f_{\pi}^2} \, I_0^T
\ ,
\nonumber \\
\end{eqnarray}
with $I_2^T = m_{\pi}^2\, I_0^T$.

\begin{figure*}
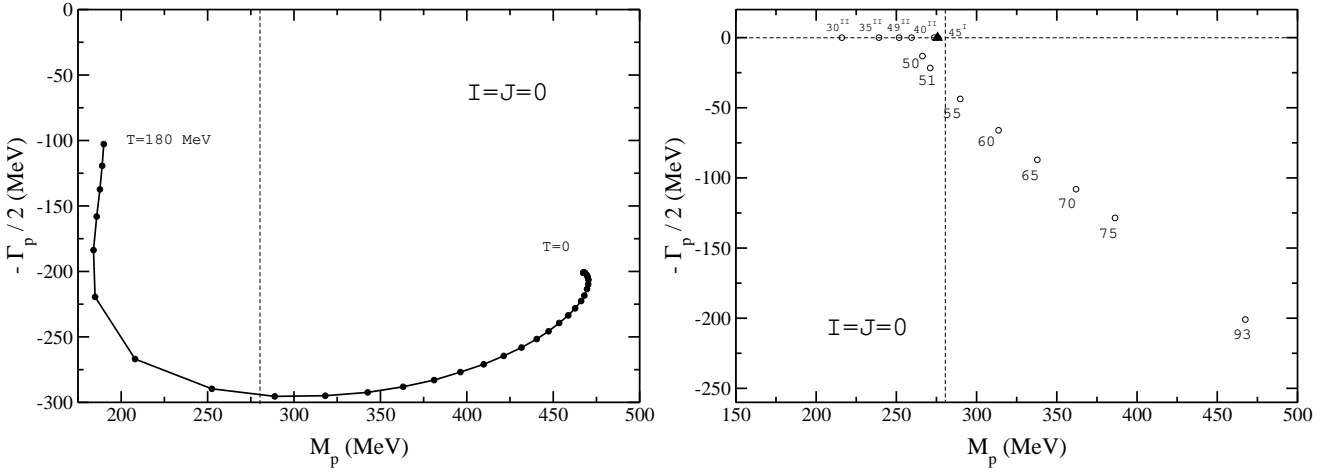

\begin{center}
\resizebox{0.48\textwidth}{!} {\includegraphics{Pole-path-LS-V2plusdV}}
\resizebox{0.48\textwidth}{!} {\includegraphics{Pole-path-LS-V2-fpiscaling}}
% Here is how to import EPS art
%\vspace{-.3cm}
%\vspace*{5cm}       % Give the correct figure height in cm
 \caption{Left: Temperature dependence of the $\sigma$ complex pole in the BS
 equation approach
 including thermal corrections to the ${\cal O}(p^2)$ kernel (the temperature
 interval between each point is 5~MeV). Right: Pole trajectory from $f_{\pi}$
 scaling. The numbers attached to each point indicate the value of $f_{\pi}$ in
 MeV.}
 \label{fig:Poles-LS}
 \end{center}
\end{figure*}

In Fig.~\ref{fig:Poles-LS} (left) we show the $\sigma$ pole trajectory for the
thermal calculation in the BS equation approach with the ${\cal O}(p^2)$ kernel
including the thermal corrections discussed above. As we can see, the evolution
of the pole follows  quite a similar pattern as in the IAM, although the
position of the $\sigma$ mass reaches values further below the two-pion threshold
for high temperatures. As in the IAM, the $\sigma$ pole stays far from the real
axis, indicating a substantial width for temperatures up to about 200~MeV,
despite the low value of $M_p$. The stronger attraction felt by the $\sigma$
meson in this approach as compared to the IAM seems to reflect that missing
thermal contributions from $t$- and $u$-channel dynamics are relevant in this
energy region and provide a repulsive contribution. Nevertheless, we obtain a
qualitatively similar physical behavior of the $\sigma$ pole in both the BS
equation approach and the IAM. Thus, we shall use the former as a starting point
to incorporate finite density effects. As a further test, we have also
calculated the $\sigma$ pole trajectory in the effective ``$f_{\pi}$ scaling''
scenario, which we depict in the right hand panel of Fig.~\ref{fig:Poles-LS}. We
obtain a similar result as in the IAM regarding the $\sigma$ pole collapsing
onto the real energy axis for $f_{\pi}$ values below approximately 50~MeV.  We
also find a pole-doubling effect which follows the ``molecular'' pattern
discussed in Sect.~\ref{sec:finitedenTzero}. Differences are observed at the
numerical level as the close-to-threshold behavior of the $\sigma$ pole is
dictated by the relative weight of the ${\cal O}(p^2)$ and ${\cal O}(p^4)$
amplitudes, which are different for the two approaches discussed in this work.
Finally, we have also studied threshold enhancement in the $\pi\pi$ amplitude
for decreasing values of $f_{\pi}$, which we omit here as our results  resemble
very much those depicted in Figs.~\ref{fig:mod00} and \ref{fig:finiteTFPI} for
the IAM calculation.

The introduction of nuclear density effects on top of the temperature
follows by a renormalization of the
pion propagator in the hot and dense medium.
In cold nuclear matter, the pion spectral
function exhibits a mixture of the pion quasi-particle mode and $p$-wave
particle-hole ($ph$), Delta-hole ($\Delta h$) excitations \cite{Oset:1981ih}.
The lowest order, irreducible $p$-wave pion self-energy
due to $ph$ and $\Delta h$ excitations reads
\begin{eqnarray}
\label{eq:piself-ph-Dh}
& &\Pi_{\pi NN^{-1}+\pi\Delta N^{-1}}^p (q_0,\vec{q};T) =
\nonumber \\
& &\left( \frac{f_N}{m_{\pi}} \right) ^2
\vec{q}\,^2 \, \left[ U_{NN^{-1}} (q_0,\vec{q};T)
+ U_{\Delta N^{-1}} (q_0,\vec{q};T) \right]
\,\,\, ,
\nonumber \\
\end{eqnarray}
where $U$ stands for the finite temperature Lindhard function, which we evaluate
in Imaginary Time Formalism (ITF) \cite{Urban:1999im,Tolos:2008di}, and the
density dependence enters $U$ through the baryon chemical potential.
We use phenomenological $\pi NN$ and $\pi N \Delta$ coupling constants
determined from analysis of pion nucleon and pion nucleus
reactions, $f_N/m_\pi = 0.007244$~MeV$^{-1}$
and $f_{\Delta}/f_N=2.13$.
The strength of the considered collective modes is modified by
repulsive, spin-isospin $NN$ and $N\Delta$ short range
correlations \cite{Oset:1981ih}, which we include in a
phenomenological way with a single Landau-Migdal interaction
parameter, $g'=0.7$. The RPA-summed (retarded) pion self-energy then reads
\begin{eqnarray}
\label{eq:piself-total}
& &\Pi^p_{\pi} (q_0,\vec{q};T) =
\nonumber \\
& &\frac{\left( \frac{f_N}{m_{\pi}} \right) ^2
F_{\pi}(\vec{q}\,^2) \, \vec{q}\,^2 \,
\left[ U_{NN^{-1}} (q_0,\vec{q};T) + U_{\Delta N^{-1}} (q_0,\vec{q};T) \right]}
{1 - \left( \frac{f_N}{m_{\pi}} \right) ^2 \, g' \,
\left[ U_{NN^{-1}} (q_0,\vec{q};T) + U_{\Delta N^{-1}} (q_0,\vec{q};T) \right]}
\,\,\, ,
\nonumber \\
\end{eqnarray}
where we have accounted for the finite size of $\pi NN$ and $\pi N
\Delta$ vertices with the hadronic monopole form factors
$F_{\pi}(\vec{q}\,^2) = \Lambda_{\pi}^2 / (\Lambda_{\pi}^2 -
\vec{q}\,^2)$, with $\Lambda_{\pi}=1300$~MeV.

\begin{figure}
\begin{center}
\resizebox{0.5\textwidth}{!} {\includegraphics{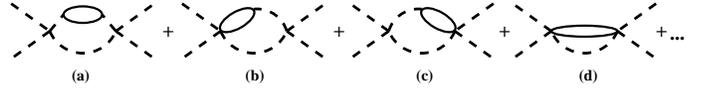}}
% Here is how to import EPS art
%\vspace{-.3cm}
%\vspace*{5cm}       % Give the correct figure height in cm
 \caption{Lowest order baryon-hole insertions in the
 meson meson loop function including vertex corrections.}
 \label{fig:diagrams2}
 \end{center}
\end{figure}

The in-medium pion propagator modifies the analytical structure of the
meson-meson loop function, $G$. At
lowest order in a density expansion (number of baryon-hole irreducible
insertions) the in-medium $\pi\pi$ amplitude at one loop includes diagram (a)
in Fig.~\ref{fig:diagrams2}, in which one of the intermediate pions
excites a $ph$ (or $\Delta h$) bubble (and similarly with the lower pion line),
on top of the vacuum $\pi\pi$ loop (Fig.~\ref{fig:diagrams2}, second
diagram).
Again one is obliged to check whether the
on-shell factorization of $V$ and $T$ is still valid in a nuclear medium. This
is actually the case for $\pi\pi$ scattering in the scalar channel as it was
shown in \cite{Chiang:1997di}.
The argument is as follows: in addition to
diagram (a), chiral symmetry requires a set of meson baryon contact terms,
depicted as diagrams (b-d) (which can be seen as contributions to the
$\pi N\to\pi\pi N$
amplitude by cutting simultaneously the $ph$ bubble and the
lower pion line in diagrams (a-d)).
It turns out that the contribution from the off-shell part of $V$ in diagram
(a) exactly cancels with the sum of the amplitudes from diagrams (b-d), leaving
us with diagram (a) with each of the $\pi\pi$ vertices factorized on-shell.
Therefore, the algebraic solution of Eq.~(\ref{eq:LS}) is no altered and one
only has
to replace the vacuum pion propagators in $G$, cf.~Eq.~(\ref{G_vacuum}),
by the in-medium ones, $D_{\pi} = [(q^0)^2-\omega_q-\Pi_{\pi}]^{-1}$.
The argument holds at finite temperature as we have checked explicitly, in
ITF, that
the same cancellation of off-shell terms takes place in the thermal amplitude.

In order to re-evaluate $G$ in the hot and dense medium we use the spectral
(Lehmann) representation of the pion propagator. The final expression for $G$,
once continued onto the real energy axis, reads
\begin{eqnarray}
\label{G_ITF}
& &G(P;\rho,T) =
\nonumber \\
& &\int_0^{\infty} \frac{d\Omega}{2\pi} \,
\left[ \frac{1}{P^0-\Omega + {\rm i}\varepsilon}
 - \frac{1}{P^0+\Omega + {\rm i}\varepsilon} \right]
 \, F(\Omega)   \ ,
 \nonumber \\
\end{eqnarray}
with
\begin{eqnarray}
\label{Fomega}
 F(\Omega) &=& \int \frac{d^3q}{(2\pi)^3}
 \int_{-\Omega}^{\Omega} du \, \pi
 \left[ 1 - n_B(E_+) - n_B(E_-)  \right]
 \nonumber \\
 &\times&
 S_{\pi}(E_+,\vec{q}\,) \, S_{\pi}(E_-,\vec{q}\,)  \ ,
\end{eqnarray}
where $S_{\pi} = -\pi^{-1} \, {\rm Im}\, D_{\pi}$ is the spectral function of the
retarded pion propagator and $E_\pm = (\Omega \pm u)/2$ (we have omitted here
the contribution from diffusion poles which typically provides a small correction
in the time-like region).
Note that ${\rm Im}\, G(P^0) = - F(P^0)/2$ and thus $F$ plays the role of a
generalized in-medium two-pion phase space including both temperature and
nuclear density effects. In the pure thermal case, $F(\sqrt{s}) =
\sigma_T(s) /8\pi\, \theta(s-4m_{\pi}^2)$.

\begin{figure}
\begin{center}
\resizebox{0.5\textwidth}{!} {\includegraphics{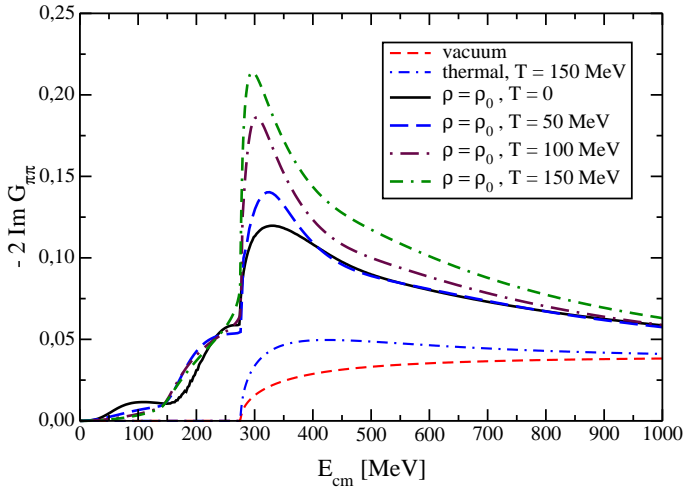}}
% Here is how to import EPS art
%\vspace{-.3cm}
%\vspace*{5cm}       % Give the correct figure height in cm
 \caption{Two-pion phase-space function  at
 finite nuclear density and temperature.}
 \label{fig:F}
 \end{center}
\end{figure}

\begin{figure}
\begin{center}
\resizebox{0.3\textwidth}{!} {\includegraphics{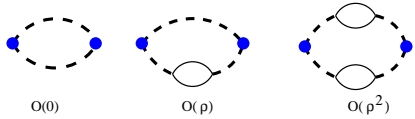}}
% Here is how to import EPS art
%\vspace{-.3cm}
%\vspace*{5cm}       % Give the correct figure height in cm
 \caption{
 ${\cal O}(1,\rho,\rho^2)$ contributions to the meson meson loop
 function from irreducible pion selfenergy insertions.}
 \label{fig:density-expansion}
 \end{center}
\end{figure}

In Fig.~\ref{fig:F} we show the phase space function, $F(P^0)$, at normal
nuclear matter density ($\rho_0=0.17$~fm$^{-3}$) and different temperatures in
the range $T=0-150$~MeV. We have also depicted the vacuum and thermal cases
for comparison. The very first difference that one observes at
finite nuclear density is the appearance of strength below the two-pion
threshold, which is absent in the thermal case. It can be understood on the
basis of the baryon-related interaction mechanisms of the pion discussed above.
At small nuclear
densities, one of the intermediate pions may excite a $ph$ pair whereas the
other is placed on the mass shell (see the second diagram in
Fig.~\ref{fig:density-expansion}).
This mechanism is responsible for the strength right below $2m_{\pi}$ and lowers
the threshold down to $m_{\pi}$. For increasing density, the probability for the two
pions to be absorbed by baryon-hole excitations sets in (third diagram in
Fig.~\ref{fig:density-expansion}), which builds additional
strength below $m_{\pi}$ and shifts
the threshold practically down to $P^0=0$ (note the smaller size of this ${\cal
O}(\rho^2)$ contribution with respect to the excitation of one single $ph$
bubble).
Beyond $P^0=2m_{\pi}$, $F(P^0)$ also exhibits a remarkable enhancement with respect
to the vacuum and thermal cases, indicating an increased phase space for $\sigma
\to \pi\pi$
decays. This reflects the widely spread structure of the pion spectral function
in the medium and the considerable attraction experienced by the pion
quasiparticle peak.

When the temperature is increased, the low energy region looses some strength,
as the phase space for $NN^{-1}$ excitations is smeared off by the thermal
motion of the nucleons. Above $P^0=m_{\pi}$, Bose enhancement is more effective
as one of the pions is placed on-shell and one can appreciate some increase over
the $T=0$ result. The latter effect is strongly magnified right above the
two-pion threshold,
since (i) the pion spectral function is populated at low energies and (ii)
the pion
quasiparticle peak is strongly attracted in the medium. Note that the rapid
energy dependence of the $p$-wave pion selfenergy is also responsible from the
quick increase of phase space right beyond the opening of the $\pi\pi$
channel.

\begin{figure*}
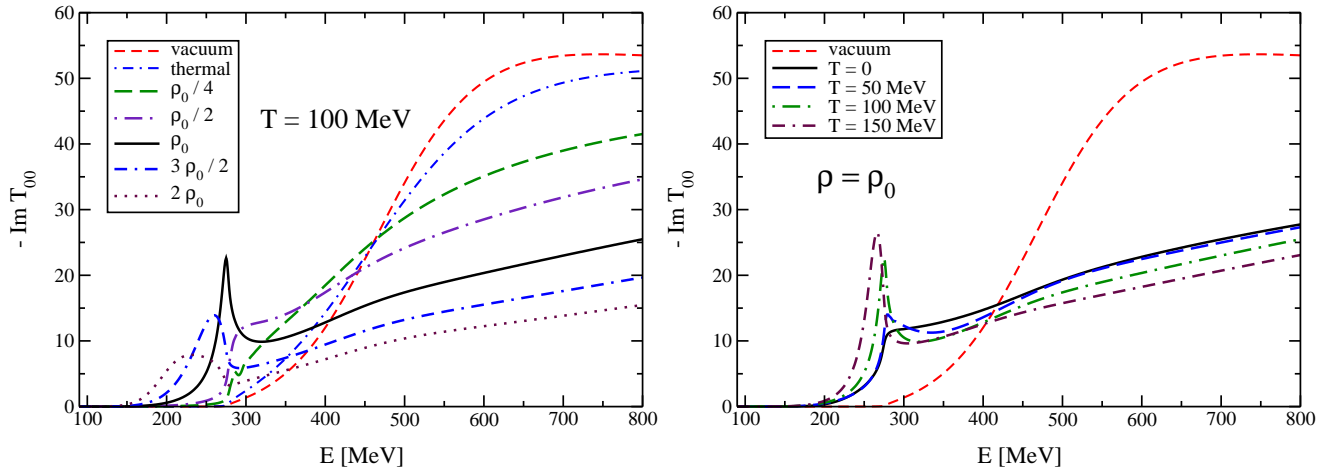

\begin{center}
\resizebox{0.48\textwidth}{!} {\includegraphics{ImT.T100.vs.rho.eps}}
\resizebox{0.48\textwidth}{!} {\includegraphics{ImT.rho0.vs.T.eps}}
% Here is how to import EPS art
%\vspace{-.3cm}
%\vspace*{5cm}       % Give the correct figure height in cm
 \caption{
 Imaginary part of the $\pi\pi$ amplitude in the $I=J=0$ channel at finite
 temperature and nuclear density. We also include the result in free space for
 reference.
 }
 \label{fig:ImT-hotdense}
 \end{center}
\end{figure*}

We finally show in Fig.~\ref{fig:ImT-hotdense} the imaginary part of the
$\pi\pi$ amplitude for different densities and temperatures, which we can
compare to the thermal calculation and the ``$f_{\pi}$ scaling'' scenario in
order to study threshold enhancement effects. We recall here that in the vacuum
and thermal cases ${\rm Im}\,T$ vanishes below the two-pion threshold. Our
result, from a dynamical calculation of many-body effects in the intermediate
two-pion state, exhibits a sizable accumulation of strength at and below
$P^0=2m_{\pi}$ already at nuclear matter density and zero temperature. On the
other hand, the amplitude is strongly depleted beyond $\sqrt{s}=400$~MeV as
compared to the vacuum case, an effect that is not shared by the
reduced-$f_{\pi}$ calculation at the equivalent density
(cf.~Fig.~\ref{fig:finiteTFPI}) which does not account for baryon-related
$\sigma$-decay mechanisms. Only for much smaller values of $f_{\pi}$ the
strength focuses at the vacuum threshold, as the $\sigma$ pole lies very close
to the real energy axis; the resulting enhancement in the amplitude in that case
is by far larger than obtained here, which is related to, but not fully driven
by the $\sigma$ pole behavior as we explain below. With respect to the thermal
calculation, finite density additionally softens the amplitude at high energies
and brings strength below the vacuum threshold, a feature which is linked to the
presence of a baryonic medium and the excitation of pionic collective modes, as
discussed above.

We have also studied the behavior of the $\sigma$ pole at finite nuclear
density. The analytic continuation of $T$ in this case is not trivial and we
have used an approximated prescription, namely to add to the evaluation of $G$
in the 1st Riemann sheet the discontinuity on the real axis with respect to the
unphysical (2nd) sheet, ${\rm Disc}\, G(P^0) = -2{\rm i} \, {\rm Im} \, G(P^0)$.
Still, this procedure provides the qualitative in-medium behavior of the
$\sigma$ pole \cite{roca02}, although its absolute position in the complex plane
carries some uncertainty (which we believe is superseded by other theoretical
uncertainties in the evaluation of the pion selfenergy). The combined effects of
temperature and nuclear density accelerate the migration of the $\sigma$ pole
towards the $\pi\pi$ threshold. For instance, at $\rho=\rho_0/2$, the $\sigma$
mass reaches $M_p=2m_{\pi}$ at about $T\simeq$150~MeV, whereas at normal nuclear
density $M_p\simeq 300$~MeV already at $T=0$ and it quickly reaches the two-pion
threshold at about $T=100$~MeV. This is correlated with the cusp structures
observed in the amplitude at threshold, cf.~Fig.~\ref{fig:ImT-hotdense};
however, the strength observed at lower energies is linked to the many-body pion
dynamics. In spite of this, the $\sigma$ remains as a broad resonance at nuclear
matter density and temperatures approaching the transition one, similarly to
what happens in the thermal calculation (at comparatively higher temperatures)
and at variance with the simplified ``$f_{\pi}$ scaling'' approximation. In
fact, if we keep increasing the density, at some point the $\sigma$ pole crosses
below $2m_{\pi}$ (one should not trust our implementation of medium effects far
beyond $\rho=\rho_0$, but as an exercise it provides information about the
phase-space behavior of the resonance). For instance, at $\rho=2\rho_0$ and
$T=100$~MeV the $\sigma$ mass from the pole lies about 70~MeV below the two-pion
threshold but still we find $\Gamma_p\simeq 150$~MeV. A more detailed
investigation of possible pole-doubling effects in this approach is on-going and
will be reported in a future work.

\section{Conclusions.}

We have presented an analysis of the behavior of $\pi\pi$ scattering amplitudes
in Unitarized Chiral Perturbation Theory with medium effects incorporated in
several ways. In particular, we have been focused on the behavior with finite
temperature and nuclear density of the $\rho(770)$ and $f_0(600)/\sigma$ resonances,
which are generated dynamically within the Inverse Amplitude Method (IAM).

By considering only thermal effects on the IAM $\pi\pi$ amplitudes, the $\rho$
exhibits a considerable broadening  with a small mass decrease as temperature
increases, whereas the $\sigma$ mainly decreases its mass, effectively signaling
chiral symmetry restoration, although it still remains as a broad resonance even
at the transition temperature. The broadening obtained in our approach for the
$\rho$ meson at finite temperature is compatible with the spectral function
analysis from dilepton spectra in the recent experiment by the NA60
Collaboration. The evolution of the $\rho$ mass with temperature does not scale
as the condensate, which renders our results in quantitative disagreement with
the Brown-Rho scaling scenario. The fact that the $\sigma$ pole remains far from
the real axis even at the two-pion threshold when only temperature effects are
considered implies no significant threshold enhancement for the scattering
amplitude, which has been advocated as a precursor of chiral symmetry
restoration. We neither observe a scaling of the $\sigma$ mass with the quark
condensate, which indicates that the $f_0(600)/\sigma$ resonance dynamically
generated in our unitarized chiral approach has a non-$\bar{q}q$ component which
is relevant near the phase transition.

By introducing finite nuclear density the picture changes
dramatically. In a first approximation we have incorporated the
effect of a nuclear medium by decreasing $f_\pi$ according to the
GOR relation to linear order in density. At sufficiently low (high)
values of $f_\pi$ (density), the $\rho$ and $\sigma$ poles collapse
onto the real energy axis at the threshold energy, which is preceded
by a significant threshold enhancement in the scattering amplitudes.
We have discussed these effects in the context of recent results
from resonance production in finite nuclei and our results are in
line with the experimental observations. A detailed analysis reveals
that when the resonance pole is close to the real axis it splits
into two states in separated Riemann sheets. The markedly different
properties of these double poles for the $\rho$ and $\sigma$
channels allows us to classify these $\pi\pi$ resonances according
to their internal structure:  whereas the $\rho$ meson presents a
clear predominant $\bar{q}q$ behavior for high densities (the two
poles stay close to threshold, one of them migrating to the 1st
Riemann sheet as a $\pi\pi$ bound state), the $\sigma$ exhibits a
``molecular'' behavior (one of the poles stays close to threshold,
well separated from the other one which evolves to lower energies to
become degenerated with the pion). The mass scaling from the
$\sigma$ and $\rho$ pole with $f_\pi(\rho)$ follows the quark
condensate evolution and therefore is compatible with a Brown-Rho
scaling scenario, although one should keep in mind that relevant
finite density mechanisms are neglected in this approximation.

Finally, we have improved our implementation of finite nuclear density (and
temperature) effects by considering a microscopic calculation of many-body pion
dynamics in $\pi\pi$ scattering. We have chosen a different unitarization scheme
for the $\pi\pi$ scattering amplitude, namely to solve the Bethe-Salpeter
equation for the lowest order ChPT interaction. Despite differences in the
amplitudes at ${\cal O}(p^4)$, this scheme essentially provides the same results
as the IAM and allows a systematic analysis and resummation of a relevant class
of pion interaction mechanisms with the nuclear medium. The pion interactions
with the medium are encoded in the single-particle pion selfenergy, which
accounts for the excitation of $p$-wave particle-hole and Delta-hole components
as well as short distance correlation effects from nucleon-nucleon and
Delta-nucleon interactions. The opening of baryon-related channels on top of
$\sigma \to \pi\pi$ at finite density extends the available phase space to lower
energies and therefore the $\pi\pi$ scattering amplitude exhibits an increased
strength at and below the two-pion threshold, which is magnified at finite
temperature as a consequence of Bose enhancement on the low energy modes of the
$\pi\pi$ intermediate states. Such an effect has been found to provide a
satisfactory description of the data from the two-pion photoproduction reaction
in nuclei when comparing the mass spectrum in the neutral- vs charged-pion
channels for different nuclei \cite{roca02,messetal}, where nuclear densities of
the order of $\rho_0$ and below are explored. In our analysis we have considered
both finite temperature and nuclear density, thus extending the applicability of
the present approach to other experimental scenarios such as the forthcoming Heavy-Ion
physics program at FAIR. As compared to the purely thermal calculation, the
attractive interaction mechanisms of the pion at finite density accelerate the
migration of the $\sigma$ pole towards the two-pion threshold. The threshold
enhancement observed in the $\pi\pi$ amplitude is correlated to the evolution of
the $\sigma$ pole towards (and below) $2m_\pi$. However, differently from the
reduced-$f_\pi$ result, the $\sigma$ pole stays far from the real axis
indicating a sizable decay width of the resonance at densities as high as
$2\rho_0$ and temperatures close to the transition one.

As a continuation of this work we plan to implement a similar many-body analysis
of finite density effects for the electromagnetic pion vector form factor, an
extension of the present work to the $SU(3)$ case where other resonances and
heavier meson states come into play, and to introduce the
effect of a finite meson-number chemical potential.
We will report on these studies elsewhere.

\section*{Acknowledgements} We acknowledge financial support from Spanish
research projects FPA2004-02602, FPA2005-02327, PR34/07-1856-BSCH, UCM-CAM
910309/08, FPA2007-29115-E and from the F.P.I.  Programme (BES-2005-6726). D.C.
wishes to thank support from the ``Juan de la Cierva'' Programme (Ministerio de
Educaci\'on y Ciencia, Spain).

%
%% For one-column wide figures use
%\begin{figure}
%% Use the relevant command for your figure-insertion program
%% to insert the figure file.
%% For example, with the option graphics use
%\resizebox{0.75\textwidth}{!}{%
%  \includegraphics{leer.eps}
%}
%% If not, use
%%\vspace{5cm}       % Give the correct figure height in cm
%\end{figure}
%%
%% For two-column wide figures use
%\begin{figure*}
%% Use the relevant command for your figure-insertion program
%% to insert the figure file. See example above.
%% If not, use
%\vspace*{5cm}       % Give the correct figure height in cm
%\end{figure*}
%\begin{table}
%\begin{tabular}{lll}
%\hline\noalign{\smallskip}
%\end{tabular}
%% Or use
%\vspace*{5cm}  % with the correct table height
%\end{table}
%
% BibTeX users please use
% \bibliographystyle{}
% \bibliography{}

\end{document}